\documentclass[ twocolumn,twocolappendix]{aastex631}
\usepackage{xspace}

\newcommand{\chandra}{\textit{Chandra}\xspace}
\newcommand{\xmmn}{\textit{XMM-Newton}\xspace}
\newcommand{\xmm}{\textit{XMM}\xspace}
\newcommand{\suzaku}{\textit{Suzaku}\xspace}
\newcommand{\nustar}{\textit{NuSTAR}\xspace}

\newcommand{\xrism}{\textit{XRISM}\xspace}

\newcommand{\kev}{\ensuremath{\;\mathrm{keV}}\xspace}
\newcommand{\kms}{\ensuremath{\;\mathrm{km\,s^{-1}}}
\newcommand{\km}{\ensuremath{\;\mathrm{km}}\xspace}
\newcommand{\pc}{\ensuremath{\;\mathrm{pc}}\xspace}
\newcommand{\kpc}{\ensuremath{\;\mathrm{kpc}}\xspace}
\newcommand{\AU}{\mbox{AU}\xspace}
\newcommand{\s}{\ensuremath{\mathrm{s}}\xspace}
\newcommand{\yr}{\ensuremath{\mathrm{yr}}\xspace}
\newcommand{\Myr}{\ensuremath{\mathrm{Myr}}}\xspace}

\newcommand{\msol}{\ensuremath{\mbox{$\mathrm{M}_{\sun}$}}\xspace}

\newcommand{\xspec}{Xspec\xspace}
\newcommand{\kt}{\ensuremath{\mathrm{kT}_\mathit{e}}\xspace}

\newcommand{\feka}{\ensuremath{\mathrm{Fe~K}\alpha}\xspace}
\newcommand{\Healpha}{\ensuremath{\mathrm{He}\alpha}\xspace}
\newcommand{\Kalpha}{\ensuremath{\mathrm{K}\alpha}\xspace}
\newcommand{\Kbeta}{\ensuremath{\mathrm{K}\beta}\xspace}
\newcommand{\Lyalpha}{\ensuremath{\mathrm{Ly}\alpha}\xspace}

\shorttitle{AASTeX v6.3.1 Sample article}
\shortauthors{Foster et al.}

\graphicspath{{./}{figures/}}

\begin{document}

\title{XMM-Newton Observations of the High Temperature Plasma in the Large Magellanic Cloud Supernova Remnant N132D}

\author[0000-0003-3462-8886]{Adam R. Foster}
\affiliation{Center for Astrophysics $|$ Harvard \& Smithsonian \\
60 Garden St.,\\
Cambridge, MA 02138, USA}

\author[0000-0003-1415-5823]{Paul P.Plucinsky}
\affiliation{Center for Astrophysics $|$ Harvard \& Smithsonian \\
60 Garden St.,\\
Cambridge, MA 02138, USA}

\author[0000-0002-5115-1533]{Terrance J. Gaetz}
\affiliation{Center for Astrophysics $|$ Harvard \& Smithsonian \\
60 Garden St.,\\
Cambridge, MA 02138, USA}

\author[0000-0003-3350-1832]{Xi Long}
\affiliation{Center for Astrophysics $|$ Harvard \& Smithsonian \\
60 Garden St.,\\
Cambridge, MA 02138, USA}
\affiliation{Department of Physics, The University of Hong Kong\\Pokfulam Road\\Hong Kong}

\author[0000-0002-1765-3329]{Diab Jerius}
\affiliation{Center for Astrophysics $|$ Harvard \& Smithsonian \\
60 Garden St.,\\
Cambridge, MA 02138, USA}

\begin{abstract}

We present an analysis of the archival XMM-Newton observations of the Large Magellanic Cloud (LMC) supernova remnant N132D totaling more than 500~ks. We focus on the high temperature plasma (\kt~$\sim4.5$\,keV) that is responsible for the high energy continuum and exciting the Fe~K emission. An image analysis shows that the Fe~K emission is mainly concentrated in the southern part of the remnant interior to the region defined by the forward shock. This Fe~K distribution would be consistent with an asymmetric distribution of the Fe ejecta and/or an asymmetric interaction between the reverse shock and the Fe ejecta. We compare the EPIC-pn and EPIC-MOS spectra in the $3.0-12.0$\,keV bandpass with a model based on RGS data plus a higher temperature component, in collisional ionization equilibrium (CIE), or non-equilibrium (NEI) (both ionizing and recombining). We find that the data are equally well-fitted by the CIE and ionizing models. %
Assuming the CIE and ionizing spectral models, the Fe in this high temperature component is significantly enhanced with respect to typical LMC abundances.  We can place only an upper limit on the neutral Fe~K line. 
We conclude that the Fe~K emission is due to ejecta heated by the reverse shock given the spatial distribution, relatively high temperature, and enhanced abundance. We estimate the progenitor mass based on the Ca/Fe and Ni/Fe mass ratios to be $13\le M_P \le 15 M_\odot$.

\end{abstract}


\section{Introduction} \label{sec:intro}

Stars more massive than 8~\msol will end their lives as core-collapse supernovae (CCSNe) or in direct collapse to a black hole  \citep{heger2003}.
CCSNe are some of the most powerful explosions in the universe,
contributing significantly to the energy budget of the interstellar medium in galaxies and 
producing a large fraction of the metals in the universe.  However, in spite of their importance, questions remain about the precise explosion mechanism.
The explosion mechanism has been the subject of intense research since the
60's \citep{bethe1990} with most efforts focused on the heating by neutrinos as the primary driver of the explosion. Recent work has emphasized the importance of multi-dimensional convection and turbulence and the breaking of spherical symmetry as key ingredients for a successful explosion \citep{janka2017J,burrows2021}. Three dimensional simulations of the explosion \citep{hammer2010,orlando2016,orlando2021,gabler2021}  indicate that convective instabilities can result in significant mixing of the innermost and outermost ejecta layers.  Of particular interest are the Fe ejecta, as they probe the regions closest to the compact object and the explosion itself.

Supernova remnants (SNRs) are complex three-dimensional objects due to asymmetries in the explosion and the surrounding medium in which the explosion occurs.
 X-ray observations of  SNRs are well-suited for studying the high temperature plasmas produced by the forward shock interacting with the circumstellar medium and the reverse shock interacting with the ejecta \citep{vink2017}.  Strong line emission from the elements of O, Ne, Mg, Si, S, Ar, Ca, \& Fe are produced by plasmas with temperatures of $10^6-10^8$~K and are accessible to the current generation of CCD instruments on the {\it Chandra X-ray Observatory} (\chandra),  \xmmn, and \suzaku.
The \feka\, line complexes between 6.4 and 7 keV can be resolved from the emission lines of other elements given the spectral resolution of the CCD instruments. \feka\, emission has been detected in young ($t<10^3~\mathrm{yr}$) and middle-aged ($t>10^3~\mathrm{yr}$) SNRs in the Galaxy and the Magellanic Clouds (MCs).
The centroid of the \feka\, complex has been used to distinguish the SNRs of CCSNe from those of Type~Ia supernovae (SNe).  \cite{yamaguchi2014} used {\textit {Suzaku}} observations to classify 23 SNRs as having a Type~Ia or core-collapse progenitor.  However, \cite{siegel2021} caution against using the centroid of the \feka\, complex derived from the entire remnant as there can be significant variations in the centroid with position in some remnants and some classifications based on the centroid disagree with the classification from other methods.

A distribution of plasmas with different temperatures, ionization histories, and abundances will contribute to the total emission of a SNR. In order to produce detectable \feka\ emission there must be plasmas present within the SNR with the appropriate conditions.
The \ion{Fe}{25} \Healpha~triplet has a peak emissivity at a temperature of $\sim6.3\times10^7$~K (or $\sim5.4$~keV) in collisional ionization equilibrium conditions, although most young and middle-aged SNRs have plasmas that are not in collisional ionization equilibrium.  
\cite{orlando2015,orlando2020} modeled the evolution of SN~1987A from supernova to supernova remnant and constructed the X-ray emission measure as function of electron temperature (\kt) and ionization timescale ($n_et$) for each of the dominant components: the shocked H~{\small II} region, the equatorial ring, and the ejecta. The emission measure distribution is broad, spanning a large range of temperatures and ionization timescales, and evolves with time. There is also significant overlap in the X-ray emission measure distribution from the three different components. An update by \cite{sapienza2023} for SN~1987A showed that the emission measure for the ejecta component peaks around temperatures of $\kt=2-3~\mathrm{keV}$ and the intensity of the ejecta component is increasing in time. \cite{sun2025} conducted a detailed differential emission measure analysis of the \xmmn\, data of SN~1987A and reached a similar conclusion that the ejecta component is increasing in time and has a temperature distribution spanning \kt$\sim1-4$~keV with a tail extending to higher temperatures.
\citet{ravi2021} fit the \chandra\, {\it High Energy Tranmission Grating} (HETG) spectra of SN~1987A with a two component thermal model and find values for the low temperature component that 
increase from 0.55~keV to 0.86~keV over a 14~year period, and values for the high temperature component which range from 1.88~keV to 2.49~keV. 
A comparable model 
\citep{orlando2021} for the Galactic SNR Cassiopeia~A (Cas~A)  showed a similar broad distribution of temperatures and ionization timescales for the plasma.  Although the distribution peaks at a temperature of around 2.0~keV, the temperatures with significant emission measure range from 0.5~keV to 5.0~keV.
Some of the ejecta knots in Cas~A have temperatures that approach or are consistent with temperatures of 5.0~keV \citep{lazendic2006,hwang2012,rutherford2013}.

Evidence for mixing of ejecta was observed in the first deep \chandra image of Cas~A \citep{hughes2000} in which Fe ejecta were found to be at larger projected radii than the ejecta of lower Z elements. A recent spectral analysis from small regions of Fe-rich ejecta features in Cas~A \citep{sato2021} detected weak features of Cr and Ti.  The authors argue that the observed Cr/Fe and Ti/Fe ratios can only be achieved by nucleosynthesis in the high-entropy nuclear burning regime (the so-called ``$\alpha$-rich freeze out zone'').  They suggest this is evidence for buoyant plumes of ejecta caused by convective instabilities in the explosion and hence supports the neutrino-driven explosion mechanism.  
Therefore studies of the \feka\, emission can
characterize the highest temperature plasmas observed in SNRs and
reveal details of the innermost ejecta shocked to these relatively high temperatures to be compared with the mixing of ejecta predicted by the 3D simulations.

The Large Magellanic Cloud (LMC) SNR N132D is the most X-ray luminous remnant in the Local Group \citep{maggi2016,sharda2020} with an X-ray luminosity of $L_\mathrm{X}(0.35-8.0~\mathrm{keV}) \sim 7.0\times10^{37}\,\mathrm{erg\,s^{-1}}$.  It was first classified as the remnant of a CCSNe \citep{danzinger1976} and an Oxygen-rich remnant \citep{lasker1978,lasker1980} based on optical observations.
\citet{blair2000} proposed that the progenitor was a massive star of $\sim25\msol$ that exploded as a  Type Ib supernova. The estimated ages range from $2500-3150\,\mathrm{yr}$
\citep{vogt2011,morse1995} based on the expansion of optical filaments. \citet{law2020} performed a 3D reconstruction of the optically-emitting O ejecta to refine the age estimate to $2450\pm195\,\mathrm{yr}$. \cite{banovetz2023}  derived a consistent age of $2770\pm500\,\mathrm{yr}$ from proper motion measurements of the O-rich ejecta based on multi-epoch observations with the {\em Hubble Space Telescope}. \citet{borkowski2007} presented the high-resolution \chandra\, image which revealed a complicated filamentary structure with a bright rim to the South and O-rich ejecta spatially coincident with optical O filaments.  
A dense CO cloud lies just to the south 
of the remnant \citep{banas1997,sano2017} and offers a possible
explanation for the enhanced X-ray emission of the
bright shell.  \citet{sano2020} presented evidence from {\it Atacama Large Milimeter Array} (ALMA) and \chandra\, that at least part of the molecular cloud complex extends to the North and is interacting with the shock of the SNR. \citet{2021arXiv210802015H} report the analysis of the {\it High Energy Stereoscopic System} (H.E.S.S.) gamma-ray spectrum which is consistent with a hadronic origin and suggest that the interaction with the molecular complex is responsible for the bright gamma-ray emission.

The explosion was suggested to have occurred in a cavity created by the strong stellar winds of the progenitor \citep{hughes1987} and such a scenario was modeled 
by \citet{chen2003}.
The forward shock velocity of the outer shell was estimated to be $\sim855\kms$ based on analysis of the X-ray spectra \citep{sharda2020}.
\citet{behar2001} analyzed the first \xmmn images and showed that the Fe~K emission was centrally concentrated inside the elliptical region defined by the lower Z elements of O, Ne, Mg, \& Si.  The \chandra\, high resolution images \citep{sharda2020} localized the Fe~K emission more precisely to the southern part of the remnant inside of the primary shell of X-ray emission. \citet{bamba2018} conducted a joint analysis of the \suzaku and {\it Nuclear Spectroscopic Telescope Array} (\nustar)  
data and found evidence for a hard component that could be fit with a plasma model close to equilibrium with a temperature of \kt$\sim5$\kev or a recombining plasma with a temperature of \kt$\sim1.5~\kev$.  They also report the first detection of the \ion{Fe}{26}\Lyalpha~line which they argue strengthens the case for a high temperature or recombining plasma.
\citet{hitomi2018} found that the \feka\, emission is red-shifted with a velocity of $\sim800~\kms$ (consistent with an ejecta origin) while the Si emission is consistent with the expected velocity of the LMC based on only 3.7~ks of high spectral resolution data from the {\it Soft X-ray Spectrometer}~\citep{kelley2016} on the {\it Hitomi} mission. 
Recent results from the {\it Resolve} calorimeter on the \xrism\, mission \citep{2024_XRISM_N132D} confirm the \feka\ is red-shifted with a value of $\sim890\pm300~\kms$.
\citet{suzuki2020} analyzed the archival \xmmn\, {\it Reflection Gratings Spectrometer} \citep[RGS,][]{denHerder2001} data  to characterize the low energy part of the spectrum from 0.3-2.0 keV to find that at least three thermal components are required to fit the data in this bandpass and that the forbidden to resonance lines ratios for \ion{O}{7} and \ion{Ne}{9} are higher than expected for a non-equilibrium, ionizing
plasma indicating that resonant scattering might be important for these lines. The RGS data are not able to provide a constraint on the high temperature plasma  as the highest temperature component required to fit these data has a value of \kt=1.36~\kev given its 0.3 to 2.1~keV bandpass.  It is clear that the X-ray emission in N132D is the result of a complex mixture of plasmas related to the forward and reverse shocks with a range of conditions (temperatures, ionization timescales, elemental abundances, etc.).  The nature and origin of a possible high temperature (\kt$\sim5$~\kev) plasma that could explain the \feka\, is not fully understood at this time.

  In this paper we anaylze the archival \xmmn\, {\it {European Photon Imaging Camera}} (EPIC) data of N132D totaling more than 500~ks in order to characterize the \feka\, emission and the high temperature plasma.  The paper is organized as follows: \S\ref{sec:data} discusses the data and the initial reduction, \S\ref{sec:image_analysis} presents the image analysis, \S\ref{sec:spectral_analysis} presents the spectral analysis, \S\ref{sec:discussion} presents our interpretation of the results and finally in \S\ref{sec:conclusions} we present our conclusions.
  We assume the distance to N132D to be $50\,\mathrm{kpc}$ in all calculations hereafter \citep{2003AJ....125.1309C,2013Natur.495...76P,2019Natur.567..200P}. At this distance, $1\arcsec = 0.24\,\mathrm{pc}$.

\section{Data and Reduction} \label{sec:data}

N132D is used as a calibration source by \xmmn, and has therefore accumulated more than 50 observations in the \xmm archive, totalling more than 1~Ms of exposure time. This provides a rich data set to examine. Due to the advanced age of the remnant ($\sim2500$ years) we do not expect the source to vary significantly during the 20 year period, and we observe no such variation during our analysis. However due to the fact that these observations were used for calibration there are a wide range of different instrument modes in use: filters, CCD modes, and aimpoints all change between observations. Some of these observations can be immediately discarded (for example, those with the filter wheel in the closed position or the $^{55}\mathrm{Fe}$ calibration source in use). Others have been chosen or discarded based on the criteria listed below for each instrument. All data were reprocessed and extracted using SAS 20.0.0 and the calibration database described by the release note XMM-CCF-REL-388, dated April 7$^{\mathrm{th}}$ 2022. Data analysis was performed using \xspec v12.12.1 \citep{1996ASPC..101...17A} through the PyXspec interface \citep{2021ascl.soft01014G}.

\subsection{EPIC PN Data Selection} \label{sec:epicpndata}

All available EPIC PN exposures, totalling 1~Ms of time, were reprocessed using the \texttt{epchain} command. We filtered the PN light curves as described in \S\ref{sec:sp_extraction} to remove times of soft proton flares. N132D is a bright source, and therefore susceptible to pileup, especially in the more sensitive PN detector. As a result, we excluded all observations not in the Prime Small Window mode, which has the fastest readout and therefore is the least affected by pileup. One consequence of this is that there is limited or no space around the remnant and the edge of the detector, as shown in Figure~\ref{fig:sampleCCDcutoffs}. This led to many observations being partially cut off. Any exposure which had less than 90\% within our remnant extraction region (see Section~\ref{sec:extraction}) and on good pixels was also excluded. After applying these criteria, all of the retained observations were 100\% on the chip.

While this paper focuses on the high energy region ($E > 4.5$\,keV), we noted slight differences in the residuals between the thin and thick PN filters at lower energies. As there were only two observations with the thick filter, we have excluded data taken with it. The final list of PN exposures, totalling 353~ks, is shown in Table~\ref{tab:pn_data}.

\subsection{EPIC MOS Data Selection} \label{sec:epicmosdata}

All available EPIC MOS exposures were reprocessed using the \texttt{emchain} command, followed by \texttt{espfilt} to remove soft proton affected data. As with the PN data, the fastest window modes were used to avoid pileup concerns. Fortunately, most of the data were in the Prime Partial W3 mode, so little data were discarded. We retain data taken with the Medium and Thin filters.  Finally, due to the increased number of bad pixels and rows in the MOS detector compared to the PN, we keep all exposures with more than 85\% of the remnant on good pixels. The selected exposures provide 775\,ks of MOS1 and 813\,ks of MOS2 data, as listed in Tables~\ref{tab:mos1_data}~and~\ref{tab:mos2_data}.

\subsection{Extraction Region}
\label{sec:extraction}
\begin{figure}
    \centering
    \includegraphics[width=8cm]{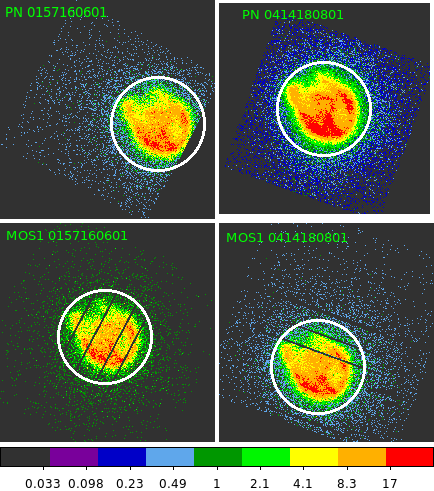}
    \caption{Sample PN PrimeSmallWindow and MOS Prime Partial mode observations, showing the 75'' radius circular extraction region used. 
    Many observations such as 0157160601 (left panels) were discarded due to the remnant being partially off the detector or too many dead pixels.\label{fig:sampleCCDcutoffs}}
\end{figure}

For the analysis presented below, we have used a 75 arcmin radius circular region  centered on RA $5^h25^m02.836^s$, Dec $-69^{\circ}38'34.86''$ encompassing the entire remnant as shown in Figure \ref{fig:sampleCCDcutoffs}. While there is some blank sky around the remnant included, N132D is so bright that the sky background contribution remains negligible, even at the high energies discussed in this work. This is fortunate as the small window mode used for most PN observations leaves little reliable sky background extraction region that we can easily make use of.

\subsection{Spectral Extraction}
\label{sec:sp_extraction}
PN and MOS Spectra were extracted using the SAS routines \texttt{evselect}, \texttt{arfgen} and \texttt{rmfgen}. To obtain the best balance of spectral accuracy and count rate at high energy, MOS data with PATTERN$\leq$12 and PN  single and double events (PATTERN $\leq4$) were accepted. High count rate times were filtered out of the MOS data using the \texttt{espfilt} histogram algorithm described in \url{https://xmm-tools.cosmos.esa.int/external/sas/current/doc/espfilt}, which fits a Gaussian to the count rate of the spectrum and marks those intervals where the rate is within 2.1$\sigma$ of the peak as good time intervals (GTI). For the PN data we had to recreate this independently as the \texttt{espfilt} routine has been disabled for Prime Small Window mode in SAS 20; our algorithm was verified by comparing with SAS 19 GTI results. Spectra from the GTIs were extracted using the standard \texttt{\#XMMEA\_EP} and \texttt{\#XMMEA\_EM} filters for the PN and MOS data respectively using the \texttt{evselect} tool.

When running \texttt{arfgen} the \texttt{applyxcaladjustment=yes} setting was set to apply the empirical cross-calibration corrections between MOS and PN\footnote{see https://xmmweb.esac.esa.int/docs/documents/CAL-SRN-0382-1-1.pdf}, as was \texttt{applyabsfluxcorr}, which adjusts the arf to make the spectral shape agree more closely with \nustar data\footnote{see https://xmmweb.esac.esa.int/docs/documents/CAL-SRN-0388-1-4.pdf}. We have also performed the analyses below with \texttt{applyabsfluxcorr=no}, and the differences encountered are not significant to this analysis: \kt changes by $<0.1$\,keV. As it does not alter any of the conclusions of this paper significantly, to avoid confusion we have only shown results for these filter settings in this paper.

\subsection{Background Modeling} \label{sec:background}

As discussed in Section ~\ref{sec:n132dfit}, the majority of our analysis will focus on the $3 < E<10$\,keV range, therefore our background estimate targets this region. The EPIC-PN observations we used were all carried out in Prime Small Window (PSW) mode.  Because of the strong variation in detector background at high energies (particularly around the Cu-K lines) it is necessary to evaluate the background for regions close to the actual position of remnant on the detector. In examining the positions in detector coordinates, it was found that most of the pointings used were around \texttt{(DETX,DETY) = (300, 1500)}.  We extracted spectra for the merged PN SW filter-wheel-closed (FWC) data set \texttt{pn\_closed\_SW\_2014\_v1.fits.gz}.  The extraction applied \texttt{FLAG==0} and included \texttt{PATTERN<=4}.  The spectra were extracted using a region specification \texttt{((DETX,DETY) in circle(293.6,1422.6,1500.0))}

A template \xspec model was constructed and fit to the FWC data.  The continuum was modeled as a power law with index 1.759 with a broad low energy gaussian (E=0.033 keV, $\sigma$=0.071 keV). The instrumental fluorescence lines were modeled with gaussians. The lines include Al-\Kalpha (1.489 keV), Au-M (2.12 keV), Ti-\Kalpha (4.51 keV), Cr-\Kalpha (5.41 keV), Fe-\Kalpha (6.4 keV), Ni-\Kalpha, Ni-\Kbeta (7.47, 8.285 keV), Cu-\Kalpha, Cu-\Kbeta (8.04, 8.905 keV), Zn-\Kalpha, Zn-\Kbeta (8.63, 9.57 keV), and Au-L$\beta$ (11.442 keV). The template was fit to the FWC data and all of the parameters were frozen except for a \texttt{const} model multiplying the model.  In the spectral fitting for the N132D data, only the \texttt{const} was free to vary: the shape of the background spectrum model was fixed but its overall normalization was allowed to vary.

For the MOS, a similar approach was followed. The model is based on the ESAS extended source model including a broken power law to capture any residual soft proton flaring emission and gaussian components for the line emission in Al-\Kalpha (1.489 keV), Si-\Kalpha (1.747keV), Cr-\Kalpha (5.41 keV), Mn-\Kalpha (5.90keV) Fe-\Kalpha (6.4 keV), Ni-\Kalpha (7.47 keV) and Au-L$\alpha$ (9.7 keV). The amplitudes of these components were obtained by fitting with filter wheel closed data in similar extraction regions to the most common detector coordinates here: centered on (DETX, DETY)=(-1350, 0) for MOS1 and (-180, -1570) for MOS2.

\section{Image Analysis} \label{sec:image_analysis}
\begin{figure*}
    \centering
    \includegraphics[width=1\textwidth]{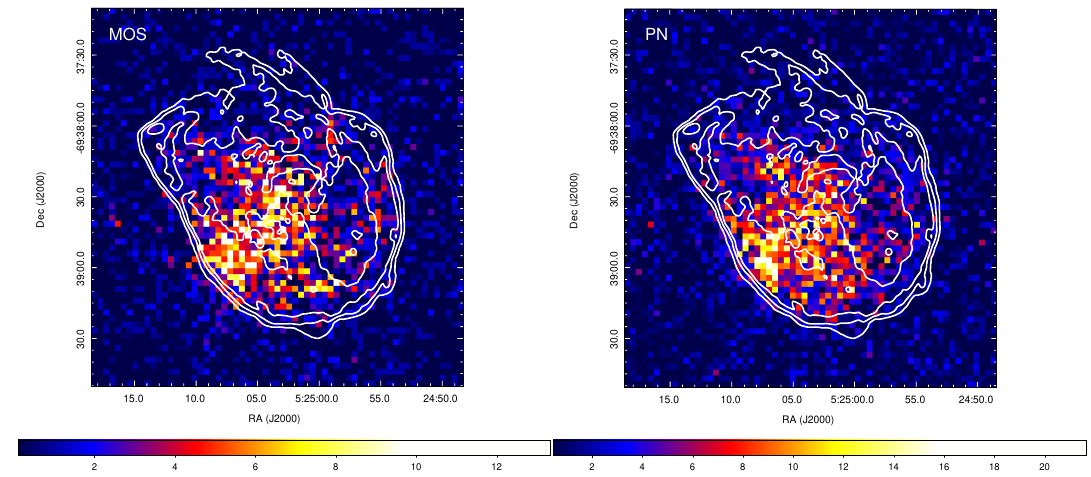}
    \caption{The Fe \Kalpha line ($6.50<E<7.05$~keV) counts image with continuum emission subtracted. The pixel size is 2$\farcs$5 $\times$ 2$\farcs$5. The contours 
    are from the (registered) and merged Chandra data (ObsIDs 5532, 7259, and 7266), in the energy band 0.5--8.0 keV.  The Fe-K emission is concentrated interior to the blast wave, and mainly in the southern part of the interior. The brightest emission is mainly central and to the east.
    \label{fig:line_img}}
\end{figure*}

We created background subtracted and exposure time corrected images for the MOS data, and exposure time corrected images for the PN data in the Fe K line band (6.50--7.05 keV) and three adjacent energy bands (5.75--6.30 keV, 6.30--6.50 keV and 7.05--7.60 keV). We estimated the continuum emission in the Fe K line region from the three adjacent energy bands. The continuum subtracted Fe K line image of N132D for the MOS and PN data are shown in Figure~\ref{fig:line_img}. The observations included in this analysis are shown in Tables~\ref{tab:pn_data}, \ref{tab:mos1_data}, $\&$ \ref{tab:mos2_data}. We reduced the EPIC-MOS data with the Science Analysis System (SAS) v20.0.0 and the Extended Source Analysis Software (XMM--ESAS) package by following the  the `cookbook for analysis procedures for XMM–Newton EPIC-MOS observations of extended objects and the diffuse background'\footnote{https://heasarc.gsfc.nasa.gov/FTP/xmm/software/xmm-esas/xmm-esas-v13.pdf}. For the EPIC-PN observations, the ESAS methods can only be used for FullFrame or Full Frame extended modes. The EPIC-PN observations we used were all carried out in small window (SW) mode. 

Therefore for the PN data, we modeled the source continuuum and detector background in the four bands as described later in this section.
We used the SAS ``image" script\footnote{https://www.cosmos.esa.int/documents/332006/641121/README.pdf} to combine the images of each energy band of all observations. 

For the MOS data, the tasks emchain and mos-filter are used to create filtered events files. The mos-spectra task is used to create the model background spectra and images, which are used in the mos\_back task to create model Quiescent Particle Background (QPB) spectra and images. The QPB spectrum is fitted in \xspec to determine the level of residual proton contamination. The proton task is used to create the model soft proton contamination maps from the \xspec results and the model soft proton detector map from mos\_back. The \texttt{comb} task is used to combine MOS1 and MOS2 images. The adapt task is used to create adaptively smoothed background subtracted and exposure corrected images. 

The background subtracted and exposure corrected images for all observations are combined for each of the four energy bands. To estimate the continuum component in the Fe K band (6.50--7.05 keV), we fit the spectrum of the remnant in the adjacent energy bands 5.75--6.30 keV, 6.30--6.5 keV and 7.05--7.60 keV with a power-law model. The flux of the power-law component in each of the four energy bands is calculated. 

The images in the three bands, 5.75--6.30 keV, 6.30--6.50 keV, and 7.05--7.60 keV were coadded, scaled by the ratio of the flux in the Fe-band to the sum of the fluxes in the other three bands, and subtracted from the Fe-band image to generate a continuum-subtracted image.
The Fe K band continuum-subtracted image for MOS data is shown in the left panel of Figure~\ref{fig:line_img}.

The ESAS method we used for the MOS data to create model background spectra and images relies on data from the unexposed (to the sky) corners of the detectors. In order to use it for the PN, the data should be in Full Frame or Full Frame Extended modes. Because the PN data we used are in Small Window mode, instead of using the ESAS method, we create exposure corrected images in the four energy bands for each observation without creating model background spectra and images. We used \texttt{epchain} and \texttt{evselect} to create filtered events files, and used \texttt{evselect} and \texttt{eexpmap} to create the exposure corrected images. The image of each observation is adaptively smoothed using \texttt{asmooth}. For each energy band the smoothed images are combined using CIAO tool \texttt{dmimgcalc}. We repeat the steps of fitting the spectrum of the remnant as in the MOS data analysis to get the background flux in the four energy bands. Instead of the power-law model, we use a power-law plus the detector background model  since we did not subtract the background of the PN data. The continuum image in the Fe-K band (6.50--7.05 keV) is obtained by co-adding the other three energy band images, and scaling by the ratio of the flux of power-law plus detector background model component in the Fe-K band to the summed fluxes of the other three energy bands. The Fe K band continuum-subtracted image for PN data is shown in the right panel of Figure~\ref{fig:line_img}. The white contours are from registered and merged 2006 Chandra data (ObsIDs 5532, 7259, and 7266) in the 0.5-8.0\,keV band.

As shown in both the MOS and PN images, the Fe-K emission is inside the blast wave and is clearly distributed differently than the bright shell of emission. The distribution of the emission is asymmetric, mainly from the center to the southeast part of the remnant with little emission in the north or southwest parts of the remnant. Though this is the distribution projected in the 2-D plane of the sky, this may suggest an asymmetric explosion. The southeast bright Fe feature is close to the edge of the remnant. This may suggest that there is dense material outside of or close to the forward shock in the southeast along the line-of-sight direction as seen in simulations of asymmetric SNe explosions \citep{wongwathanarat2015,orlando2021}. These results are similar to the Chandra observations of \citet{sharda2020}, where the \feka\ emission is reported to be in the interior of the southern half of the remnant, and not extending out to the blast wave.

\section{Spectral Analysis} \label{sec:spectral_analysis}

The N132D spectra were selected and extracted as described in Section \ref{sec:data}. Spectral analyses were performed using \xspec. We model the instrumental background separately for the PN and each MOS, using the method described in Section~\ref{sec:background}. In all the analyses described in this section, each of the 83 spectra are fit simultaneously with the same model of N132D. For plotting purposes only, we have summed all the PN spectra and/or all the MOS spectra and re-binned the spectra to provide a 10 sigma significance in each bin. All analysis was done with unbinned spectra and fits were optimized using the C statistic \citep{cash1979}. As the C statistic per degrees of freedom is not a reliable goodness of fit metric, we have included the Pearson $\chi^2_P$ in the later fits. All uncertainties reported in this paper are the $1\sigma$ uncertainties unless noted otherwise.

\subsection{Low Energy Spectrum \texorpdfstring{$(0.3 < E < 2.0$}{} keV)}
\label{sec:suzuki}

For the lower energy spectrum, we adopt the model of \cite{suzuki2020}. This is a three \texttt{bvrnei} model, assuming the same initial temperature (0.01\,keV) and ionization timescale $(9.8\times10^{10}$\,cm$^{-3}$ s), but three different final temperatures (0.2, 0.563 and 1.36\,keV). We applied this model and froze all parameters except for the overall normalization, and then fit the spectrum. In all instruments, this model matches the overall normalization of the emission in the $0.3 < E < 2.0$\,keV region, but it over predicts the emission above $E> 2$\,keV. We therefore uncoupled the normalization of the 1.36\,keV \texttt{bvrnei} component from the others, and an approximately acceptable fit for the low energy region was found. The resulting overall normalization of this fit is $5.08\times10^{12}\,\mathrm{cm}^{-5}$, a 16\% increase on the value published by \citet{suzuki2020} of $4.38\times10^{12}\,\mathrm{cm}^{-5}$, while the norm of the 1.36~keV component (obtained as explained in section \ref{sec:n132dfit}) is 0.029, compared with 0.031 in the Suzuki paper. Combining the two norm factors, the overall increase in the 1.36~keV component emission measure is $\approx~9\%$ between the Suzuki model and ours. As a reminder, the \xspec\ norm is in units of $10^{14}\,\mathrm{cm}^{-5}$, so the overall normalization is a ``norm'' of 0.0508 in \xspec.

\begin{figure}
    \centering
    \includegraphics[width=3.2in]{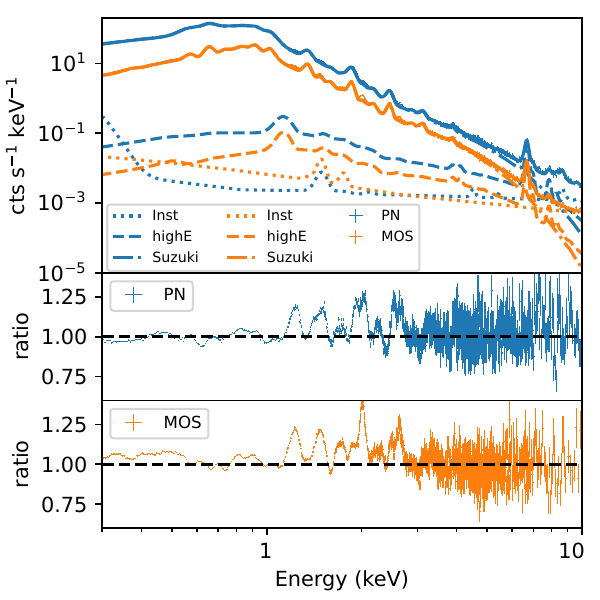}
    \caption{The joint equilibrium fit to the MOS and PN data using the model described in Section~\ref{sec:n132dfit}. Shown are: the total spectrum (solid lines); the high temperature component (dashed lines); the Suzuki model (dash-dot lines) and the instrumental background (dotted lines). PN data are in blue and combined MOS 1 and 2 data are in orange. } \label{fig:wholefit}
\end{figure}

As shown in Table \ref{tab:suzuki_norms} and Figure \ref{fig:wholefit} the fits are not statistically acceptable, however they do capture the overall amplitude of the N132D flux. Jointly fitting PN and MOS data in the $1.5 < E < 3$\,keV region is difficult due to differences between the two instrument's spectra (see Figure~\ref{fig:wholefit}) which cannot be resolved by changing the model, and we therefore do not attempt this here, instead focusing on the $E > 3$\,keV region.

To do this, we froze the norm for the Suzuki model, excluding that of the $kT=1.36$\,keV component as it was the only part which provided significant emission in the $E>2.0$\,keV range. This provides a model which, when combined with our additional components for the the high energy region, both avoids over predicting the low energy spectrum and reasonably represents the contributions of the lower temperature components to the high energy spectrum.

\begin{deluxetable}{lll}
    \caption{The normalizations obtained from fitting the \citet{suzuki2020} model to the MOS and PN spectra in the $0.3 < E < 10.0$\,keV range.\label{tab:suzuki_norms}}
\tablehead{
\colhead{Instrument}        &\colhead{Norm ($\times10^{12} \,\mathrm{cm}^{5}$)} &  \colhead{cstat/dof}   }
\startdata
PN     & $5.040^{+0.002}_{-0.002}$ & 58669.2/5812 \\
MOS    & $5.127^{+0.002}_{-0.002}$ & 121112.6/23728 \\
ALL    & $5.077^{+0.001}_{-0.001}$ & 184305.9/29542 \\
\enddata
\end{deluxetable}

\subsection{Instrumental Background Level}
\label{sec:background_fit}
N132D is a bright source such that instrumental background is only relevant when the effective area of the \xmm\ instruments drops at high energy. It is about 10\% of the signal at $E=6$~keV and becomes the dominant contribution at $E\gtrapprox 7.2$\,keV. As we are focusing on the Fe \Healpha complex, accurately estimating the continuum level can have significant effects on the resulting fits in this region. 

\begin{figure}
    \centering
    \includegraphics[width=3.2in]{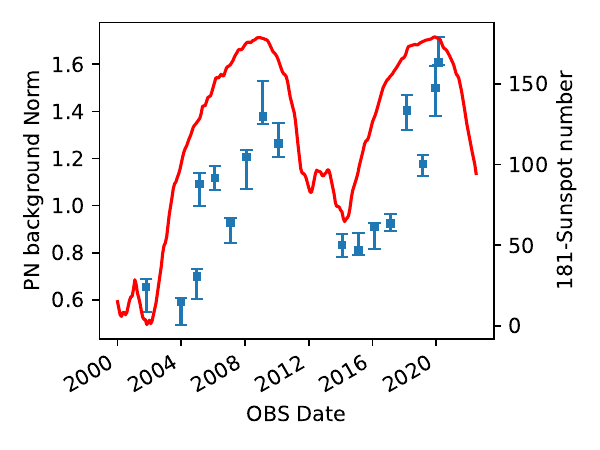}
    \caption{Normalization of the PN background (blue squares) compared with the observed number of sunspots (red line; SILSO data, Royal Observatory of Belgium, Brussels). \label{fig:pn_background}}
\end{figure}

The many different observations of N132D used here, particularly for the PN in small window mode, make it difficult to extract pure background spectra with sufficient counts to be useful. We adopted the approach of specifying the shape of the instrument background spectrum as described in Section~\ref{sec:background} while allowing the overall normalization to vary to account for spatial variations on the detector and temporal variations of the incident particle background.  Initially, we attempted to use three single normalizations, one for each of MOS1, MOS2 and PN. However, the PN data have sufficient counts at high energy that  statistically significant differences in the incident particle spectrum are detected throughout the solar cycle, which varies considerably over the 20 year period in which these data were collected (see Figure~\ref{fig:pn_background}). We therefore freed each background normalization for the 17 PN spectra, while maintaining global MOS1 and MOS2 background normalizations.

The spectrum between $6.0 < E < 12.0$ was fitted with an initial approximation of our high energy model (see Section~\ref{sec:n132dfit}), the Suzuki model (frozen) and the instrumental background models. As the instrument background models are dominant above $E=8$\,keV, their normalization is set by the emission in this region. 

\subsection{High Energy Spectrum \texorpdfstring{$(3.0 < E < 12.0$ keV)}{}}
\label{sec:n132dfit}

The high energy spectrum includes features from the Ar-\Kalpha lines at $E\approx3.1$\,keV, Ca-\Kalpha lines at $E\approx3.9$\,keV, the Fe-\Kalpha complex at $E\approx6.7$\,keV, and the Ni-\Kalpha lines at $E\approx7.8$\,keV. The Suzuki model, as fitted in Section \ref{sec:suzuki}, models the Ar lines well without modification. However it contains very little Ca (abundance is 0.05 times solar), therefore the Ca lines are underpredicted. As it's highest temperature component is $\kt=1.36$\,keV, it also provides only a small inner shell ($E\approx6.55$\,keV) contribution to the Fe-\Kalpha complex and no significant emission in the Ni-\Kalpha band. There is therefore clearly a hotter plasma component contributing to the strong Fe-\Kalpha emission.

We have modeled this high energy contribution as \texttt{TBvarabs*TBvarabs*(bvvrnei+gaussian+powerlaw)}. The \texttt{TBvarabs} components are fixed to the values from \citet{suzuki2020} and represent galactic (solar abundances from \citealt{Wilms2000}, $n_H=6.2\times 10^{20}$\,cm$^{-2}$) and LMC absorption (abundances from \citealt{1992ApJ...384..508R,2016AJ....151..161S}, $n_H=6.8\times 10^{20}$\,cm$^{-2}$). The \texttt{bvvrnei} represents the additional high temperature component required to create the Fe-K emission, with the abundances of Ca, Fe and Ni free. An extra Gaussian is included, fixed at 6.4keV, in case there is additional neutral Fe-\Kalpha flux as reported by \citet{bamba2018}. The power law represents the cosmic X-ray background, with a fixed $\Gamma=1.4$ and $\mathit{norm}=4.56\times10^{-6}$ $\mathrm{photons}\,\mathrm{keV}^{-1}\,\mathrm{cm}^{-2}\, \mathrm{s}^{-1}$ at $E_\mathit{phot}=1\,\mathrm{keV}$ based on the area of our 75'' radius extraction region and the results of \citet{2004A&A...419..837D}. Finally, given the overlap in the possible source of the Ca emission - both the $\kt=1.36$\,keV and hot component could contribute - we have freed the calcium abundance in the 1.36\,keV component of the Suzuki model.

\begin{figure}
    \centering
    \includegraphics[width=3.2in]{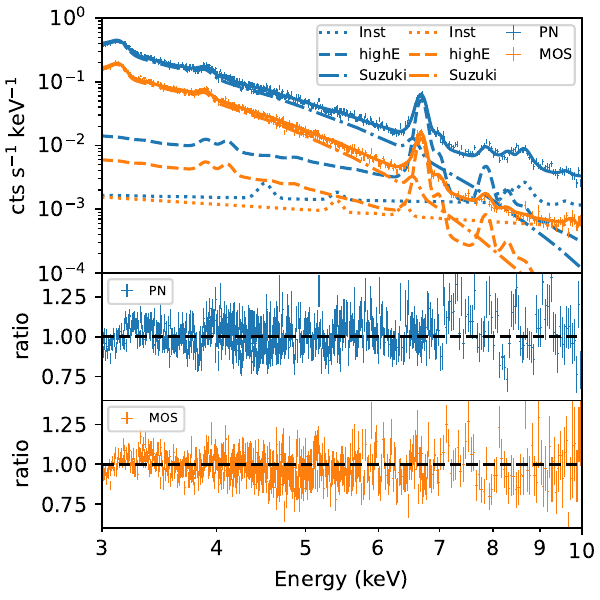}
    \includegraphics[width=3.2in]{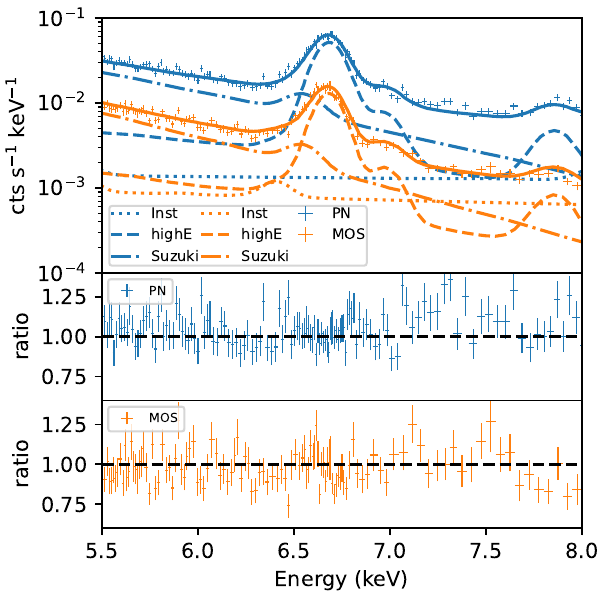}
    \caption{The same as Figure~\ref{fig:wholefit}, but zoomed in on the  (top) $3<E<11$~keV and (bottom) $5.5<E<7.5$~keV range. \label{fig:wholefit_equilibrium_3_10keV}}
\end{figure}

Due to the complexity of the model and the tendency to find local minima, we fit these data using Monte-Carlo methods via the the \xspec chain command with 100,000 iterations of burn in and then 300,000 iterations for the chain itself. Figures~\ref{fig:wholefit} and \ref{fig:wholefit_equilibrium_3_10keV} show the resulting fit for the equilibrium case for several energy ranges.

\begin{deluxetable}{llll}
    \caption{Joint fits using PN and MOS for the $3<E<10$ keV. The Ca$_{\mathrm{Suz}}$ and norm$_{\mathrm{Suz}}$ refer to the $kT=1.36$~keV component of the \citet{suzuki2020} model. For comparison, LMC ISM abundances are Ca=0.49, Fe=0.26, Ni=0.98 \citep{1992ApJ...384..508R, 2016AJ....151..161S}.
\label{tab:fitting_results_n132d_Ca}}
\tablehead{\colhead{Component}   & \colhead{Equilibrium} & \colhead{Ionizing} & \colhead{Recombining}   }
\startdata
kT\tablenotemark{a}   & $4.49^{+0.13}_{-0.09}$  & $9.85^{+0.09}_{-0.22}$  & $4.50^{+0.10}_{-0.23}$ \\
Ca\tablenotemark{b}                    & $8.64^{+3.31}_{-2.37}$  & $11.49^{+1.70}_{-2.62}$  & $9.15^{+3.70}_{-1.63}$ \\
Fe\tablenotemark{b}                    & $7.21^{+0.77}_{-0.41}$  & $7.95^{+0.36}_{-0.70}$  & $7.13^{+0.76}_{-0.49}$ \\
Ni\tablenotemark{b}                    & $8.51^{+3.23}_{-1.72}$  & $5.54^{+1.52}_{-2.64}$  & $8.67^{+2.63}_{-0.57}$ \\
norm\tablenotemark{c}                  & $3.31^{+0.20}_{-0.34}$  & $1.92^{+0.19}_{-0.07}$  & $3.33^{+0.32}_{-0.31}$ \\
Ca$_{\mathrm{Suz}}$\tablenotemark{b}   & $0.94^{+0.03}_{-0.05}$  & $0.86^{+0.05}_{-0.06}$  & $0.93^{+0.03}_{-0.06}$ \\
norm$_{\mathrm{Suz}}$\tablenotemark{d} & $2.96^{+0.02}_{-0.01}$  & $3.00^{+0.01}_{-0.01}$  & $2.96^{+0.02}_{-0.01}$ \\
$\tau$\tablenotemark{e}                  & N/A                     & $0.25^{+0.01}_{-0.01}$  & $1.76^{+0.07}_{-1.00}$ \\
  d.o.f. & 154792 & 154791 & 154791\\
  cstat & 118754 & 118755 & 118754\\
  $\chi^2_P$ & 156493 & 155364 & 156358\\
\enddata
\tablenotetext{a}{in keV} \tablenotetext{b}{in interstellar medium values \citep{Wilms2000}} \tablenotetext{c}{in $10^{10}$cm$^{-5}$} \tablenotetext{d}{in $10^{12}$cm$^{-5}$} \tablenotetext{e}{in $10^{12}$ cm$^{-3}$s}
\end{deluxetable}

\begin{deluxetable}{llll}
    \caption{Same as Table~\ref{tab:fitting_results_n132d_Ca}, but showing results for MOS and PN separately in the equilibrium case.\label{tab:fitting_results_n132d_Ca_byinst}}
\tablehead{\colhead{Component}   & \colhead{Joint} & \colhead{PN} & \colhead{MOS}   }
\startdata
kT\tablenotemark{a}                       & $4.49^{+0.13}_{-0.09}$  & $4.54^{+0.17}_{-0.10}$  & $4.46^{+0.14}_{-0.19}$ \\
Ca\tablenotemark{b}                       & $8.64^{+3.31}_{-2.37}$  & $14.31^{+4.04}_{-3.06}$  & $6.47^{+1.83}_{-2.70}$ \\
Fe\tablenotemark{b}                       & $7.21^{+0.77}_{-0.41}$  & $5.31^{+0.59}_{-0.37}$  & $10.06^{+1.98}_{-1.34}$ \\
Ni\tablenotemark{b}                       & $8.51^{+3.23}_{-1.72}$  & $7.92^{+1.55}_{-1.46}$  & $6.00^{+5.83}_{-3.25}$ \\
norm\tablenotemark{c}                     & $3.31^{+0.20}_{-0.34}$  & $4.55^{+0.31}_{-0.50}$  & $2.33^{+0.37}_{-0.39}$ \\
Ca$_{\mathrm{Suz}}$\tablenotemark{b}      & $0.94^{+0.03}_{-0.05}$  & $0.81^{+0.08}_{-0.08}$  & $0.98^{+0.05}_{-0.05}$ \\
norm$_{\mathrm{Suz}}$\tablenotemark{d}    & $2.96^{+0.02}_{-0.01}$  & $2.91^{+0.03}_{-0.01}$  & $3.00^{+0.02}_{-0.02}$ \\
  d.o.f.                        & 154792 & 30592 & 124192\\
  cstat                         & 118754 & 30507 & 88178\\
    $\chi^2_P$ & 156493 & 30797 & 124493\\
\enddata
\tablenotetext{a}{in keV} \tablenotetext{b}{in interstellar medium values \citep{Wilms2000}} \tablenotetext{c}{in $10^{10}$cm$^{-5}$} \tablenotetext{d}{in $10^{12}$cm$^{-5}$}
\end{deluxetable}
Table~\ref{tab:fitting_results_n132d_Ca} shows the results of fitting in this energy band for three proposed plasma scenarios: equilibrium,  ionizing, and recombining. Some limitations were placed on the fits: the upper temperature of the ionizing plasma was limited to 10\,keV, and the timescale $\tau$ of the ionizing and recombining plasma were limited to $2\times 10^{12}$\,cm$^{-3}$\,s. The fit statistics of the three cases are essentially indistinguishable. However, for the ionizing plasma, the preferred temperature is above 6\,keV, while for the recombining plasma, the timescale approaches the upper limit, at which point the plasma is almost in equilibrium; we have relaxed the timescale further but found no significant change in our results.  

In all three cases, some commonalities arise: there is a component with $\kt > 4$\,keV, which is iron rich (7 to 8 times ISM); this component is about 1-2\% of the normalization of the Suzuki component; and there is no requirement for an additional 6.4\,keV line component beyond that provided by inner shell ionization from the 1.36~keV Suzuki component, which contains significant($>10$\%) Fe$^{18-21+}$. The contribution can be seen in the lower part of Figure~\ref{fig:wholefit_equilibrium_3_10keV}. This potential 6.4\,keV component is not listed in Table~\ref{tab:fitting_results_n132d_Ca}, but in all cases only an upper limit of order $3\times 10^{-7}$\,cm$^{-2}$\,s$^{-1}$ was obtained. Calcium is robustly detected in the $\kt=1.36$\,keV component, with a $>3\sigma$ detection in the hot component. The total quantity of Ca in the hot component is consistent across all three models -- $\mathit{norm}\times \mathrm{Ca} \approx 25$. Nickel is detected at a similar significance to Calcium. In all cases, the abundances of Ca, Fe and Ni are elevated from typical LMC values.

The temperature for the ionizing fit is substantially higher than that of the equilibrium fit. This is due to the fact that the continuum of this component is never the dominant one, with the $\kt=1.36$\,keV component dominating below $E<8$\,keV, and the background dominating above. As a result, this component is modeling the H- and He-like Fe line ratios while being almost unconstrained by the shape of the continuum, and the higher temperature of the ionizing component ensures that, despite the low ionization timescale, the ion fraction is similar for the three cases: Fe$^{23+}: 7 (11,8)\%$, Fe$^{24+}: 70 (73,70)\%$ and Fe$^{25+}: 20 (15,19)\%$ for the equilibrium (ionizing, recombining) plasma respectively. Due to their indistinguishability, throughout the remainder of our analysis, we will focus on the equilibrium results.

\subsection{MOS vs PN results}
\label{subsection:mos_pn}
We compared equilibrium model fits for the MOS and PN independently. These are shown in Table~\ref{tab:fitting_results_n132d_Ca_byinst}. The normalizations of the hot component vary significantly between the instruments; however the total mass of iron (abundance $\times$ normalization) is
consistent between the instruments. The PN favors more Ca in the hot component, reflecting the centroid of the line being at a slightly higher energy in the PN spectrum ($3.915\pm0.012$\,keV for PN, $3.896\pm0.005$\,keV for MOS)\footnote {The statistical uncertainties in the fit are 1\,eV for MOS and 2\,eV for PN; we have quoted the  systematic gain uncertainties from the MOS and PN (see \url{https://xmmweb.esac.esa.int/docs/documents/CAL-TN-0018.pdf}) of 5\,eV and 12\,eV respectively, which dominate the uncertainties.}; the rest energy of resonance line is 3.902\,keV \citep{10.1063/1.5121413}. The MOS detection of Ca and Ni is of weaker significance than the PN, with Ni $<2\sigma$. This can also be seen by eye in the Figure~\ref{fig:wholefit}, where there is no obvious feature in the MOS spectrum at $E=7.8$\,keV.

\subsection{K Shell Line Centroids}
\label{subsection:centroids}
In the high energy spectrum, the Ca, Fe and Ni $n=2\rightarrow 1$ emission can be seen clearly (except for Ni in MOS) at $E=$3.9, 6.7, and 7.8\,keV respectively. This emission could be caused by the He-like triplet, or by inner shell lines of lower charge states of Fe. For an ionizing plasma, inner shell excitation or ionization can create a hole in the $1s$ shell, which if filled by a $2p\rightarrow1s$ electron decay emits a \Kalpha\ photon. The energy centroid of the \feka\ line gradually increases as the ionization state increases, as shown in Figure 3 of \citet{2014ApJ...780..136Y}. Similarly such states can be produced by dielectronic recombination, where resonant inner shell excitation occurs to capture a free electron leaving the $1s$ vacancy, although this is suppressed in ionization dominated plasma. If we know $T_e$ from the continuum emission fits, we can therefore tell if a plasma has reached ionization equilibrium by measuring the centroid of the K line emission and comparing that with the expected dominant ionization stage(s) at the electron temperature in equilibrium. 

\begin{figure}
    \centering
    \includegraphics[width=3.2in]{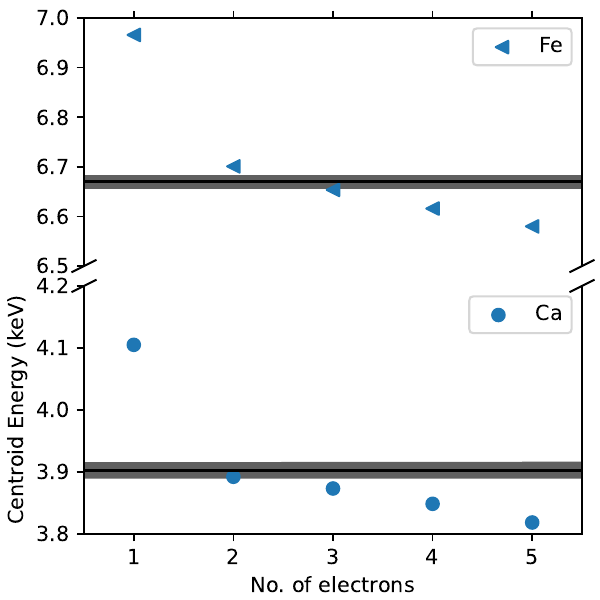}
    \caption{The centroids of the $n=2\rightarrow 1$ emission for Ca and Fe ions. The observed centroid from Gaussian fitting is shown as a horizontal black line. Shaded regions show the $\pm 12$\,eV systematic uncertainty from the PN detector. \label{fig:centroids}}
\end{figure}

To attempt to assess whether these lines come from equilibrium or non-equilibrium emission, we have made fits using the frozen best fit results for the equilibrium case above, but with the Fe and Ca abundance set to zero and replaced with a single Gaussian (Ca) or two Gaussians (one with a free energy for the He-like Fe-K emission, one with a fixed energy of the Fe \Lyalpha).  For Ca, a joint PN and MOS data fit was made between $3.5 \le E \le 4.3$~keV, and $5.5 \le E \le 7.5$~keV for Fe, resulting  in centroids with $3.903^{+0.002}_{-0.003}$ and $6.670^{+0.001}_{-0.001}$~keV respectively.\footnote{Again, we note that these statistical uncertainties are substantially smaller than the gain uncertainty of the EPIC instruments.} The observed line centroids correspond to lines of Ca$^{18+}$ and a combination of Fe$^{23+}$ and Fe$^{24+}$ respectively, as shown in Figure~\ref{fig:centroids}. These results for Fe do not rule in or out non-equilibrium but they do place upper limits on the time scale for a non-equilibrium plasma (otherwise the centroid would drift further towards Fe$^{24+}$) and the temperature of the equilibrium case. In addition, the small/absent Ca \Lyalpha feature at $E=4.1$~keV places a limit on the fraction of the Ca emission being from the same hotter plasma as the Fe \Healpha, whether in equilibrium or not.

An alternative interpretation, similar to the conclusions of \citet{hitomi2018}, is that there is a slight red-shift to the equilibrium He-like Fe emission, of $7\pm 1$\,eV, or $310\pm 45\, \mathrm{km}\,\mathrm{s}^{-1}$, while the Ca emission would be slightly blue-shifted by $5$\,eV. However given the resolution of the EPIC detectors these purely statistical error bars are likely to be unrepresentatively narrow and therefore we do not claim any measurement of redshift. Similary, the recent {\em{X-ray Imaging and Spectroscopy Mission}} \xrism observations of N132D \citep{2024_XRISM_N132D} show that there is a difference in line broadening between the He-like and H-like Fe emission lines ($51\pm 3$\,eV and $21\pm5$\,eV respectively). With the lower resolution of the EPIC cameras, we obtain for the broadenings $31^{+12}_{-29}$\,eV and $58^{+63}_{-58}$\,eV respectively, with only a very small change to the fit statistic ($\Delta$cstat$=8$, cstat$=33960$) compared with line widths frozen at zero. We are therefore unable to put more than upper limits on the broadening, which are consistent with the results from \citet{2024_XRISM_N132D}.

\subsection{He-like Fe Fits}
\begin{figure}
    \centering
    \includegraphics[width=3.2in]{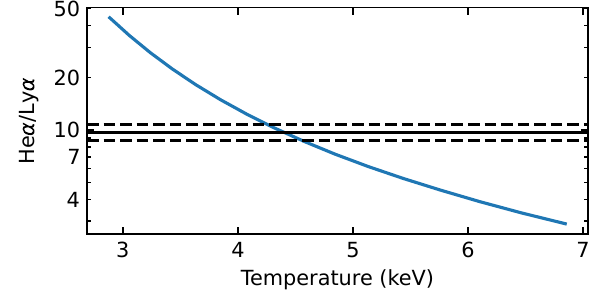}
    \includegraphics[width=3.2in]{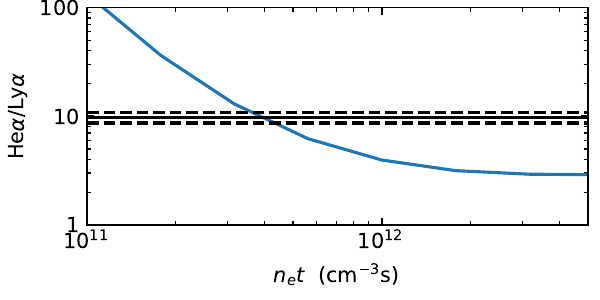}
    \caption{ The theoretical Fe-K \Healpha/\Lyalpha ratio from AtomDB for (top) an equilibrium plasma and (bottom) an ionizing plasma with $kT=6.82\,\mathrm{keV}$. The uncertainty range obtained from the values in Table~\ref{tab:fitting_gaussians_n132d} is indicated by the horizontal lines. The equilibrium temperature range is $4.25 < kT < 4.57\,\mathrm{keV}$ or the implied ionization timescale is $3.65 < n_et < 4.33 \times 10^{11}\,\mathrm{cm}^{-3}\,\mathrm{s}$. \label{fig:lyaratio}}
\end{figure}

The above discussion combines the flux from the Suzuki model at lower temperature with our additional component at high temperature. To analyze the Fe \Healpha complex in detail we have also modeled this region as a simple series of Gaussians. We used 5 gaussians, with fixed energies at 6.4\,keV (to represent inner shell \Kalpha emission), 6.637, 6.675 and 6.700\,keV (the He-like Fe forbidden, intercombination and resonance lines) and at 6.966\,keV (the H-like Fe\Lyalpha~line). There were modeled on top of a continuum thermal model with all the line emission removed, \texttt{nlapec}, and the frozen Suzuki model from Section~\ref{sec:n132dfit}. The fit was performed between 5.5 and 7.5\,keV. Due to the low resolution of the spectrum, the intercombination line amplitude was fixed to 65\% of the resonance line, based on typical values from the AtomDB database for equilibrium plasma in the $1<kT_e<10$\,keV range. The fit results are shown in Table~\ref{tab:fitting_gaussians_n132d}, and plotted against the theoretical values in Figure~\ref{fig:lyaratio}.

The continuum temperature obtained from this fit, $\kt=6.82$~keV, is slightly higher than that implied from the line ratios in equilibrium ($4.40 < \kt < 4.62$~keV). If we fix the temperature at $\kt=6.82$~keV and compare with line ratios for an ionizing plasma (see bottom panel of Figure~\ref{fig:lyaratio}), we obtain an ionization timescale of $3.96^{+0.37}_{-0.31}\times 10^{11}$\,cm$^{-3}$\,s, which is slightly higher than is implied by the ionizing case of the joint fits ($2.53^{+0.12}_{-0.13}\times 10^{11}$~cm$^{-3}$s).  However, the ion fraction for Fe remains similar to the results from Section~\ref{sec:n132dfit}: Fe$^{23+}$:$6\%$, Fe$^{24+}$:$74\%$ and Fe$^{25+}$:$18\%$. This reflects that while there is significant degeneracy in the relative contribution of the spectral components to the continuum, the ion fraction is quite stable between these models.

\begin{deluxetable}{llll}
    \caption{The results of joint fits using PN and MOS for the $5.5<E<7.5$ keV spectrum with a \texttt{nlapec} plus 5 gaussian model.  \label{tab:fitting_gaussians_n132d}}
\tablehead{
\colhead{Component}   &\colhead{Value}}
\startdata
\kt  & $6.82^{-0.37}_{-1.4}$  keV\\
norm$_\mathrm{nlapec}$ & $3.69^{+0.69}_{-0.07}$  $\times 10^{10}$ cm$^{-5}$\\
norm$_\mathrm{6.4\,keV}$ & $<7.91$  $\times 10^{-8}$ cm$^{-2}$s$^{-1}$\\
norm$_\mathrm{He\,f}$ & $3.45^{+0.32}_{-0.35}$  $\times 10^{-6}$  cm$^{-2}$\,s$^{-1}$\\
norm$_\mathrm{He\,r}$ & $6.13^{+0.27}_{-0.19}$  $\times 10^{-6}$  cm$^{-2}$\,s$^{-1}$\\
norm$_{\mathrm{Ly}\alpha}$ & $1.30^{+0.13}_{-0.09}$ $\times 10^{-6}\,\mathrm{ cm^{-2}\,s^{-1}}$\\
d.o.f.   & 34707\\
cstat    & 33960\\
\enddata
\end{deluxetable}

\section{Discussion} \label{sec:discussion}

\subsection{High Temperature Plasma and Fe Ejecta Mass}
\label{sec:highTplasma}

The presence of a high temperature plasma ($\kt=4.5-7.0$~keV) in N132D was indicated previously by \xmmn~\citep{behar2001}, \chandra~\citep{sharda2020}, and \nustar~\citep{bamba2018}, and is confirmed by our spectral fits to the combined archival \xmmn data.  The high resolution spectrum from the {\em{Resolve}} calorimeter instrument   on \xrism~\citep{2024_XRISM_N132D} confirms the existence of such a high-temperature plasma and cleanly resolves the \ion{Fe}{25}~\Healpha line complex from the \ion{Fe}{26}~\Lyalpha line.

\citet{suzuki2020} modeled N132D using a spherically symmetric shell structure with an outer radius $R \sim 1\arcmin$ and a thickness of $R/12$. If we assume that this is true for the the high temperature component, then we estimate a total hydrogen mass of 21 $M_{\odot}$, or a total mass of $29\,M_{\odot}$ if all elements are included (principally He) with a  density of $n_H=0.30, n_e=0.36, n_{Fe}=5.8\times10^{-5}$\,cm$^{-3}$. However, the continuum-subtracted images of the Fe-K emission shown in Figure~\ref{fig:line_img} indicate that this emission is concentrated in the central and southeastern parts of the remnant. We can encompass nearly all of this emission in a circle of radius $r=37''$, and  model this high temperature component as coming entirely from a uniform sphere of that radius centered at RA=$5^h25^m05^s$, Dec=$-69^\circ38'45''$. This provides a similar mass estimate, $M_H=22.7 M_{\odot}, M=31.6 M_{\odot}, M_{Fe}=0.24\pm0.02\,M_{\odot}$, however it should be noted that this is an upper limit on mass (and a lower limit on density): it assumes a uniform filling factor of 1, as opposed to any clumping or density gradient within the sphere. Any such clumping would lead to higher local densities, and a similar amount of total emission from a lower total mass of material. As both geometries give similar results, to avoid repetition we will focus on the spherical case from here on.

An electron density of $n_e=0.36$\,cm$^{-3}$, combined with the age of the remnant of approximately 2700 years (see Section \ref{sec:intro}) gives an ionization timescale, $\tau=n_e t = 3\times10^{10}$\,cm$^{-3}$s, far below the equilibrium requirement of $\tau\gtrapprox 10^{12}$\,cm$^{-3}$s, and lower than the ionizing plasma case result of $2.5\times10^{11}$\,cm$^{-3}$s. To achieve this $\tau$ in 2700 years, we require $n_e\gtrapprox3$\,cm$^{-3}$.

The XMM MOS and PN instruments have a a Half Energy Width of $\approx 15$'' on axis\footnote{\url{https://xmm-tools.cosmos.esa.int/external/xmm_user_support/documentation/uhb/basics.html}}. This implies that the emitting region could be significantly smaller than the 37'' radius assumed, even before accounting for clumping of the emitting material. With the fixed observed Fe line intensity, which is proportional to:
\begin{equation}
\label{eqn:volume}
I \propto Vn_en_{Fe} \propto Vn_e^2\frac{n_{Fe}}{n_e}
\end{equation}
and since $\tau \propto n_e$, $\tau \propto r^{-3/2}$ , where $r$ is the radius of a uniform sphere of emitting material and $V$ is its volume. We can therefore also obtain the same observed Fe line intensity and the required $n_e=3.0$ to achieve $\tau=2.5\times10^{11}$\,cm$^{-3}$s if instead the emission comes from a smaller, more dense pockets within this sphere. A uniform density sphere of $r=9.3''$ or uniformly dense clumps with a filling factor of 1.5\% within the 37'' sphere would satisfy these requirements.  This scenario would give a Fe mass of $M_{Fe}=0.030\pm0.003\,M_{\odot}$.

We postulate that the high temperature component consists mainly of shocked ejecta based on its temperature, abundances and spatial distribution. The enhanced abundance of metals present -- approximately 10 times larger than the surrounding ISM values -- cannot be explained by the surrounding media, strongly implying that the source is ejecta. The central location of the emission also lends itself to an ejecta origin, as there is no significant emission outside the shock front. Some portion nearest the shock front in the southeast could plausibly be ascribed to the ISM based on the location, but it is a modest fraction and does remain inside the shock front. The directions with the strongest \feka\ emission are those in the southern half of of the remnant, where N132D is expanding into a molecular cloud and therefore the reverse shock is most advanced, raising the temperatures of this shocked ejecta. This contrasts to the north of the remnant, where there is no high density cloud and therefore expansion continues freely or with less deceleration, with little or no reverse shock forming, and therefore much of the ejecta in this direction may remain invisible to us.

\citet{2024_XRISM_N132D} proposed a pure metal ejecta plasma as a possible source of the emission for the H- and He-like Fe emission. Equation~\ref{eqn:volume}, and the $n_e=3\,\mathrm{cm}^{-3}$ requirement from the timescale, allows us to explore  the change from $n_{Fe}/n_e$ from 1/6190 in an H and He dominated plasma to $\approx 1/40$ (all metals, i.e. no H or He, with abundances from our model) or $\approx 1/24$ (for a pure Fe plasma, dominated by Fe$^{24+}$). These imply that equivalent uniform spheres of 1.7'' (all metal) and 1.45'' (pure iron) and $M_{Fe}=0.030\pm0.003M_\odot$ could produce the same emissivity and timescales. Alternatively, this can be viewed as clumpy material with a filling factor of 0.01\% within the 37'' sphere.

The progenitor models presented in \cite{2016ApJ...821...38S} give a range of iron yields of between 0.07 and 0.14\,$M_{\odot}$ for progenitors between $12.25$ and $27.3\,M_{\odot}$. This Fe mass is also comparable to the $0.13\,M_{\odot}$ of Fe ejecta found in \chandra observations of the well-studied Galactic SNR Cassiopeia~A~\citep{hwang2012}.  Our observed numbers of between $0.030 < M_{Fe} < 0.24\,M_{\odot}$ cover this range, but with a large uncertainty. %

We note: (1) our assumption of uniform density is an oversimplification, and (2) we can only see the shocked ejecta, whereas nucleosynthesis models account for all of the iron. Therefore we base our analysis on the ratio of the elemental abundances, as these will be less susceptible to absolute uncertainties. 

\subsection[The 6.4 keV Feature]{Possible Neutral Fe Feature}

There is a small contribution to the \ion{Fe}{25}~\Healpha line complex shown in Figure~\ref{fig:wholefit_equilibrium_3_10keV} from inner shell emission of Fe due to the kT$\sim 1.36$~keV component of the Suzuki model. The centroid of this complex is at $E=6.55$\,keV. In addition there is a Fe \Kalpha contribution from the MOS background. 
If the $\kt=1.36$~keV component is instead replaced with a noline apec model, thereby keeping the continuum shape but removing the $E=6.55$\,keV feature, the norm of the  $E=6.4$\,keV gaussian becomes $1.82^{+0.99}_{-1.09}\times10^{-7} \,\mathrm{cm}^{-2}\,\mathrm{s}^{-1}$. This is approximately~3\% of the resonance line intensity, so it is weak, but present. We note that the results presented in \cite{2024_XRISM_N132D} did not include a neutral Fe-K line.
Therefore we can confirm through this fitting that there is a feature due to inner shell ionization, but we cannot distinguish between inner shell ionization of Fe L-shell ions and neutral iron fluorescence. Given that the ionizing $kT=1.36$\,keV component models the other spectral features, we suggest that the Fe L-shell ions are the more likely origin.

\subsection{Ca and Ni Abundances}
\label{sec:CaAbund}
As described in Section~\ref{sec:n132dfit}, we freed the Ca abundance in the $\kt=1.36$~keV component of the Suzuki model to account for the flux at 3.9~keV. The original Suzuki model has a Ca abundance of 0.05 Ca$_{\mathrm{ISM}}$ for all components, which becomes approximately 0.9 Ca$_{\mathrm{ISM}}$ in our fits described above. If we instead keep the Suzuki abundance of 0.05 Ca$_{\mathrm{ISM}}$ and free the Ca abundance of the high temperature component, the abundance becomes $20^{+3}_{-1}$ Ca$_{\mathrm{ISM}}$, although the fit is slightly worse (cstat=120065 in the equilibrium case, compared with 118754 with the Suzuki component free, see Table~\ref{tab:fitting_results_n132d_Ca}) due to the \Lyalpha feature being over-predicted. 

If we assume that the calcium component of the high temperature component has the same spatial origin as the proposed 37'' radius sphere proposed for the Fe in section~\ref{sec:highTplasma}, we obtain $M_{\mathrm{Ca}}=0.012^{+0.005}_{-0.003}M_{\odot}$ for this component. Making the assumption that this high temperature component represents ejecta, and that the Ca follows a similar spatial distribution to the Fe, we can compare the mass ratios in this component, giving $M_{\mathrm{Ca}}/M_{\mathrm{Fe}}$ = $0.051^{+0.020}_{-0.015}$. 

The nickel abundance is more difficult to account for, as it is dependent on the instrument used - PN detects it robustly, MOS does not. This reflects that the expected Ni signal at 7.8keV is barely above the instrument background and continuum combination for the MOS, and the general uncertainty in the continuum shapes. If we follow the same assumptions as for Ca above, assuming the PN data to be correct, we obtain $M_{\mathrm{Ni}}=0.012^{+0.005}_{-0.003}\,M_{\odot}$, with $M_{\mathrm{Ni}}/M_{\mathrm{Fe}}$ ratio of $0.052^{+0.020}_{-0.012}$. 

In both cases, the numbers given here are for the equilibrium case, the ionizing plasma is similar and therefore omitted for simplicity. If the emission comes from smaller regions, as explored in Section~\ref{sec:highTplasma}, the ratios for these remain the same even if the absolute masses vary, under the assumption that the spatial distribution of the three metals is similar.

\subsection{Progenitor Mass}
\label{sec:progenitormass}

\begin{figure}
    \centering
    \includegraphics[width=3.2in]{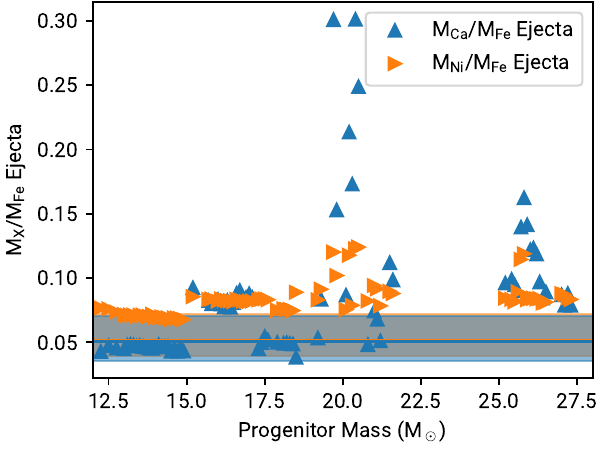}
    \caption{The Ca/Fe (blue) and Ni/Fe (orange) ejecta mass ratios from the W18 model of \citet{2016ApJ...821...38S}. The shaded area indicates the results from this work, which largely overlap and therefore appear grey.} \label{fig:ejecta}
\end{figure}

If we interpret the hottest component of Fe, Ca and Ni emission as shocked ejecta, while the Suzuki components are swept up ISM, we can estimate the mass of the progenitor ($M_P$) of N132D using these abundance ratios.  %
In Figure~\ref{fig:ejecta} we compare our results to the ejecta yields for the W18 CCSN model of \citet{2016ApJ...821...38S}. We note this model is for core collapse of a non-rotating single stars with a solar elemental abundance; we use it here to check the plausibility of our results. From this figure, acceptable progentior mass range is $13\le M_P\le 15\,M_{\odot}$. This broadly agrees with estimates from \textit{Chandra},  $M_P=15\pm 5M_\odot$ \citep{sharda2020}, and \nustar and \suzaku, $M_P=15M_\odot$ \citep{2018ApJ...863..127K}. It is, however, substantially lower than the estimates of $30 - 35 M_{\odot}$ \citep{blair2000} and $50^{+25}_{-15} M_{\odot}$ \citep{2009ApJ...707L..27F} taken from UV/optical and \textit{Hubble} observations.

\section{Conclusions}
\label{sec:conclusions}
We have analyzed a very extensive but diverse set of N132D calibration observations from the \textit{XMM-Newton} EPIC cameras to study the emission in the  $E\ge 3$\,keV band. 
We confirm the existence of a high temperature plasma $(\kt=4.49\pm0.13\,\mathrm{keV})$ assuming a CIE model with enhanced abundances of Ca, Fe, \& Ni that are $\approx10\times$ typical LMC abundances. Spectral fits with the PN and MOS instruments separately agree on the temperature and enhanced abundances, but disagree on the degree of enhancement.
This high temperature thermal component is required to obtain the observed Fe line ratios, and is preferred for the Ca ratios. A similar component was identified and modeled as a recombining plasma based on \nustar observations \citep{bamba2018}. We find that the recombining plasma is slightly less favored than an ionizing or equilibrium plasma. We are unable to meaningfully distinguish between the latter two with the resolution of the EPIC cameras. Comparing Fe \Healpha\ and \Lyalpha line ratios independently with these data produces similar results, suggesting an ionizing or equilibrium plasma with $\kt \ga 4$\,keV.

We constructed continuum-subtracted images of the Fe-K emission and compared them to the low-energy X-ray emission that defines the extent of the remnant.  The Fe-K emission is located interior to the region defined by the forward shock, mostly in the southern part of the remnant. The brightest Fe-K emission is located in the center and southeastern parts exhibiting a different morphology than the low-energy X-ray emission. The asymmetric spatial distribution of the Fe-K and enhanced abundances for the high temperature component are consistent with an ejecta origin for this Fe emission.

Observations with \nustar and \textit{Hitomi} have suggested the presence of a neutral Fe \Kalpha line at 6.4\,keV \citep{bamba2018, hitomi2018}, however this line is not significantly detected in our work. That said,  \citet{bamba2018} give the luminosity of the neutral line as $1.9\times10^{33}$\,erg\, s$^-1$, which corresponds to a normalization of $6.72\times 10^{-7}$\,cm$^{-2}$\,s$^{-1}$. The upper limits we are observing are of order $3 \times 10^{-7}$\,cm$^{-2}$\,s$^{-1}$.

We have, under the assumption that the abundances of the hot component are representative of ejecta heated by the reverse shock, estimated the progenitor mass based on the Ca/Fe and Ni/Fe mass ratios to be $13\le M_P \le 15 M_\odot$, based on models from \citet{2016ApJ...821...38S}. This broadly agrees with estimates from other X-ray observations, which indicate an intermediate mass progenitor, while being lower than those from UV/optical observations. We note, however, that this is dependent on the shocked ejecta which we can observe being representative of all the remnant's ejecta.

\begin{acknowledgments}

\begin{center}
{\bf Acknowledgements}
\end{center}

This work was supported by NASA Astrophysics Data Analysis Program grant NNX11AD17G and NASA XRISM Grant 80NSSC23K1656.  P.P.P., T.J.G. \& D.J. acknowledge support from the Smithsonian Institution and the Chandra X-ray Center through NASA contract NAS8-03060. X.L. is supported by a GRF grant of the Hong Kong Government under HKU 17304524.

We are grateful for many useful discussions over the years on this analysis with members of the {\it International Astronomical Consortium for High Energy Calibration} (IACHEC) including Steven Sembay, Eric Miller, Brian Grefenstette, Martin Stuhlinger, Andrew Beardmore, Konrad Dennerl, and Andrew Pollock.\\

\end{acknowledgments}

\vspace{5mm}
\facilities{XMM}

\appendix

\label{sec:app_data}

In this appendix we provide details of all the observations which were used in this work.
\begin{deluxetable*}{lrrrllrr}
    \caption{The XMM EPIC PN data used in this analysis. DETX and DETY refer to the location of the center of the remnant on the detector, in detector coordinates. \% Remnant refers to the fraction of the circular extraction area which is on the CCD chip, excluding bad pixels. For the PN, data were excluded if they were not in Prime Small Window mode, their exposure was $<10$ks, their filter was not Thin1 or Medium, or $<90\%$ of the remnant was included. \label{tab:pn_data}}

\tablehead{
\colhead{Exposure}        &\colhead{Filter} &  \colhead{DETX}   & \colhead{DETY} &  \colhead{DATE}       &  \colhead{MODE}           & \colhead{Exposure (ks)} & \colhead{\% Remnant}}
\startdata
0129340901 pnS001     & Medium & 585.7  &1676.3&  2001-10-20 &  PrimeSmallWindow & 15.7   &100.0\%\\
0129341401 pnS022     & Thin1 &  359.7  &1491.1&  2003-12-24 &  PrimeSmallWindow &17.0   &100.0\%\\
0129341701 pnS022     & Thin1 &  335.7  &1560.3&  2004-12-21 &  PrimeSmallWindow &16.5   &100.0\%\\
0129341801 pnS022	  & Medium & 306.5  &1429.4&  2005-02-19 &  PrimeSmallWindow & 17.5    &100.0\%\\
0129342001 pnS022     & Thin1 &  306.0  &1401.4&  2006-02-07 &  PrimeSmallWindow &19.3   &100.0\%\\
0414180101 pnS001     & Thin1 &  309.0  &1408.9&  2007-02-05 &  PrimeSmallWindow &21.7   &100.0\%\\
0414180201 pnU002     & Thin1 &  300.8  &1426.9&  2008-02-06 &  PrimeSmallWindow &17.1   &100.0\%\\
0414180401 pnS001     & Thin1 &  310.1  &1424.6&  2009-02-12 &  PrimeSmallWindow &12.4   &100.0\%\\
0414180501 pnS001     & Thin1 &  359.0  &1426.1&  2010-02-08 &  PrimeSmallWindow &22.3   &100.0\%\\
0414180601 pnS001     & Thin1 &  321.5  &1436.3&  2014-02-05 &  PrimeSmallWindow &32.9   &100.0\%\\
0414180701 pnS001     & Thin1 &  323.4  &1436.7&  2015-02-12 &  PrimeSmallWindow &23.6   &100.0\%\\
0414180801 pnS001     & Thin1 &  309.1  &1412.7&  2016-02-15 &  PrimeSmallWindow &30.3   &100.0\%\\
0414180901 pnS001     & Thin1 &  295.2  &1433.3&  2017-02-17 &  PrimeSmallWindow &24.5   &100.0\%\\
0811012401 pnS001     & Thin1 &  294.6  &1387.4&  2018-02-23 &  PrimeSmallWindow &17.6   &100.0\%\\
0811012501 pnS001     & Thin1 &  276.9  &1377.5&  2019-02-27 &  PrimeSmallWindow &24.6   &100.0\%\\
0853782401 pnS001     & Thin1 &  702.2  &1437.9&  2019-12-11 &  PrimeSmallWindow &12.7   &100.0\%\\
0811012601 pnS001     & Thin1 &  315.6  &1412.0&  2020-02-23 &  PrimeSmallWindow &28.5   &100.0\%\\
                      &       &         &      &             &  \textbf{Total}   &353 & \\
\enddata

\end{deluxetable*}

\begin{deluxetable*}{lrrrllrr}
    \caption{As table \ref{tab:pn_data}, for the XMM EPIC MOS1 data used in this analysis. For MOS, data were excluded if they were not in Prime Partial W3 mode, their exposure was $<5$ks, their filter was not Thin1 or Medium, or $<85\%$ of the remnant was not included. \label{tab:mos1_data}}

\tablehead{
\colhead{Exposure}        &\colhead{Filter} &  \colhead{DETX}   & \colhead{DETY} &  \colhead{DATE}       &  \colhead{MODE}           & \colhead{Exposure (ks)} & \colhead{\% Remnant}}
\startdata
0125100201 mos1U009   &Medium & -259.9  &-250.7&  2000-05-23 &  PrimePartialW3   &12.2   &100.0\%\\
0125100301 mos1S002   &Medium & -254.3  &-244.3&  2000-05-23 &  PrimePartialW3   & 6.9   &100.0\%\\
0129340601 mos1S002   &Medium &  926.1  & 283.6&  2000-09-29 &  PrimePartialW3   &17.7   &97.8\%\\
0129340801 mos1S002   &Thin1  & -1260.8 &-896.9&  2001-06-25 &  PrimePartialW3   &19.0   &99.9\%\\
0129340901 mos1S002   &Medium & -1602.1 & 276.8&  2001-10-20 &  PrimePartialW3   &22.2   &98.6\%\\
0129341101 mos1S014   &Medium &  355.6  &-202.4&  2002-07-11 &  PrimePartialW3   &13.9   &99.9\%\\
0157160301 mos1S003   &Medium &  126.8  &-124.6&  2002-11-10 &  PrimePartialW3   &25.7   &91.5\%\\
0157160601 mos1S003   &Medium &  73.1   &-119.0&  2002-11-16 &  PrimePartialW3   &22.4   &90.1\%\\
0157160801 mos1S003   &Medium &  10.0   &-176.6&  2002-11-24 &  PrimePartialW3   &25.4   &88.5\%\\
0157161001 mos1S003   &Medium &  74.6   &-214.2&  2002-12-14 &  PrimePartialW3   &29.9   &88.9\%\\
0157360201 mos1S003   &Medium &  74.6   &-197.8&  2002-12-31 &  PrimePartialW3   &14.1   &93.6\%\\
0157360301 mos1S003   &Medium &  103.9  &-212.4&  2003-01-17 &  PrimePartialW3   &27.7   &93.0\%\\
0157360501 mos1S003   &Medium &  222.7  &-273.3&  2003-02-22 &  PrimePartialW3   & 9.9   &95.5\%\\
0129341401 mos1S023   &Medium & -1417.8 &  50.1&  2003-12-24 &  PrimePartialW3   &24.1   &95.5\%\\
0129341501 mos1S023   &Thin1  & -1340.7 &  56.9&  2004-02-24 &  PrimePartialW3   &25.0   &97.8\%\\
0129341601 mos1S023   &Thin1  & -1260.5 & 138.3&  2004-06-17 &  PrimePartialW3   &16.8   &95.1\%\\
0137551101 mos1S002   &Thin1  &  229.2  &-150.0&  2004-09-08 &  PrimePartialW3   &12.3   &93.0\%\\
0210681301 mos1S004   &Medium & -1579.1 &  49.2&  2004-11-25 &  PrimePartialW3   &10.8   &97.5\%\\
0129341701 mos1S023   &Medium & -1487.1 &  26.3&  2004-12-21 &  PrimePartialW3   &24.0   &97.4\%\\
0129341801 mos1S023   &Thin1  & -1356.3 &  -3.4&  2005-02-19 &  PrimePartialW3   &20.0   &95.7\%\\
0129342001 mos1S023   &Thin1  & -1328.4 &  -4.0&  2006-02-07 &  PrimePartialW3   &26.1   &95.8\%\\
0414180101 mos1S002   &Thin1  & -1335.9 &  -0.9&  2007-02-05 &  PrimePartialW3   &31.1   &95.3\%\\
0414180201 mos1U002   &Thin1  & -1353.8 &  -9.1&  2008-02-06 &  PrimePartialW3   &21.7   &98.0\%\\
0414180401 mos1S002   &Thin1  & -1351.5 &   0.2&  2009-02-12 &  PrimePartialW3   &14.2   &93.1\%\\
0414180501 mos1S002   &Thin1  & -1352.9 &  49.2&  2010-02-08 &  PrimePartialW3   &32.3   &95.3\%\\
0414180601 mos1S002   &Thin1  & -1363.2 &  11.6&  2014-02-05 &  PrimePartialW3   &47.4   &95.4\%\\
0414180701 mos1S002   &Thin1  & -1363.5 &  13.6&  2015-02-12 &  PrimePartialW3   &32.6   &95.6\%\\
0414180801 mos1S002   &Thin1  & -1339.6 &  -0.8&  2016-02-15 &  PrimePartialW3   &41.3   &95.4\%\\
0414180901 mos1S002   &Thin1  & -1360.2 & -14.7&  2017-02-17 &  PrimePartialW3   &35.2   &92.7\%\\
0811012401 mos1S002   &Thin1  & -1314.4 & -15.4&  2018-02-23 &  PrimePartialW3   &23.0   &95.4\%\\
0811012501 mos1S002   &Thin1  & -1304.6 & -33.2&  2019-02-27 &  PrimePartialW3   &37.2   &95.3\%\\
0853782401 mos1S002   &Thin1  & -1363.3 & 392.4&  2019-12-11 &  PrimePartialW3   &16.6   &90.7\%\\
0811012601 mos1S002   &Thin1  & -1338.9 &   5.6&  2020-02-23 &  PrimePartialW3   &36.4   &94.7\%\\
                      &       &         &      &             &  \textbf{Total}   &775 & \\
\enddata
\end{deluxetable*}

\begin{deluxetable*}{lrrrllrr}
    \caption{As table \ref{tab:pn_data}, for the XMM EPIC MOS2 data used in this analysis. For MOS, data were excluded if they were not in Prime Partial W3 mode, their exposure was $<5$ks, their filter was not Thin1 or Medium, or $<85\%$ of the remnant was not included. \label{tab:mos2_data}}

\tablehead{
\colhead{Exposure}        &\colhead{Filter} &  \colhead{DETX}   & \colhead{DETY} &  \colhead{DATE}       &  \colhead{MODE}           & \colhead{Exposure (ks)} & \colhead{\% Remnant}}
\startdata
0125100201 mos2U009   &Medium &    79.0 & -477.4& 2000-05-23 &  PrimePartialW3   &12.4   &94.1\%\\
0125100301 mos2S003   &Medium &    72.6 & -471.8& 2000-05-23 &  PrimePartialW3   & 7.0   &94.1\%\\
0129340601 mos2S003   &Medium &  -449.2 &  711.3& 2000-09-29 &  PrimePartialW3   &17.5   &93.1\%\\
0129340801 mos2S003   &Thin1  &   720.1 &-1481.6& 2001-06-25 &  PrimePartialW3   &19.7   &96.7\%\\
0129340901 mos2S003   &Medium &  -455.3 &-1816.9& 2001-10-20 &  PrimePartialW3   &22.8   &93.9\%\\
0129341101 mos2S003   &Medium &    33.9 &  138.3& 2002-07-11 &  PrimePartialW3   & 9.4   &93.9\%\\
0157160301 mos2S004   &Medium &   -45.1 &  -90.0& 2002-11-10 &  PrimePartialW3   &25.9   &93.8\%\\
0157160601 mos2S004   &Medium &   -51.0 & -143.8& 2002-11-16 &  PrimePartialW3   &22.5   &93.8\%\\
0157160801 mos2S004   &Medium &     6.3 & -207.1& 2002-11-24 &  PrimePartialW3   &25.8   &93.9\%\\
0157161001 mos2S004   &Medium &    44.2 & -142.7& 2002-12-14 &  PrimePartialW3   &30.1   &94.0\%\\
0157360201 mos2S004   &Medium &    27.8 & -142.6& 2002-12-31 &  PrimePartialW3   &14.7   &94.0\%\\
0157360301 mos2S004   &Medium &    42.5 & -113.4& 2003-01-17 &  PrimePartialW3   &27.2   &94.0\%\\
0157360501 mos2S004   &Medium &   104.0 &    5.0& 2003-02-22 &  PrimePartialW3   &10.9   &94.0\%\\
0129341301 mos2S024   &Medium &   -34.1 &    8.2& 2003-09-13 &  PrimePartialW3   &17.0   &93.8\%\\
0129341401 mos2S024   &Medium &  -227.7 &-1633.8& 2003-12-24 &  PrimePartialW3   &23.9   &91.2\%\\
0129341501 mos2S024   &Thin1  &  -234.1 &-1556.7& 2004-02-24 &  PrimePartialW3   &24.9   &91.2\%\\
0129341601 mos2S024   &Thin1  &  -315.1 &-1476.0& 2004-06-17 &  PrimePartialW3   &17.3   &91.1\%\\
0137551101 mos2S003   &Thin1  &   -19.2 &   12.2& 2004-09-08 &  PrimePartialW3   &13.0   &91.1\%\\
0210681301 mos2S007   &Medium &  -227.7 &-1795.1& 2004-11-25 &  PrimePartialW3   &10.4   &91.3\%\\
0129341701 mos2S024   &Medium &  -204.3 &-1703.2& 2004-12-21 &  PrimePartialW3   &23.8   &91.3\%\\
0129341801 mos2S024   &Thin1  &  -173.9 &-1572.5& 2005-02-19 &  PrimePartialW3   &23.0   &91.2\%\\
0129342001 mos2S024   &Thin1  &  -173.2 &-1544.6& 2006-02-07 &  PrimePartialW3   &26.7   &91.3\%\\
0414180101 mos2S003   &Thin1  &  -176.3 &-1552.1& 2007-02-05 &  PrimePartialW3   &31.6   &91.3\%\\
0414180201 mos2U002   &Thin1  &  -168.2 &-1570.1& 2008-02-06 &  PrimePartialW3   &21.2   &91.3\%\\
0414180401 mos2S003   &Thin1  &  -177.5 &-1567.7& 2009-02-12 &  PrimePartialW3   &15.9   &91.3\%\\
0414180501 mos2S003   &Thin1  &  -226.4 &-1568.8& 2010-02-08 &  PrimePartialW3   &32.0   &91.2\%\\
0414180601 mos2S003   &Thin1  &  -189.0 &-1579.3& 2014-02-05 &  PrimePartialW3   &47.6   &91.3\%\\
0414180701 mos2S003   &Thin1  &  -190.9 &-1579.7& 2015-02-12 &  PrimePartialW3   &34.5   &89.2\%\\
0414180801 mos2S003   &Thin1  &  -176.4 &-1555.8& 2016-02-15 &  PrimePartialW3   &42.0   &89.2\%\\
0414180901 mos2S003   &Thin1  &  -162.6 &-1576.5& 2017-02-17 &  PrimePartialW3   &35.0   &89.2\%\\
0811012401 mos2S003   &Thin1  &  -161.7 &-1530.7& 2018-02-23 &  PrimePartialW3   &27.2   &89.2\%\\
0811012501 mos2S003   &Thin1  &  -143.9 &-1521.0& 2019-02-27 &  PrimePartialW3   &38.7   &89.2\%\\
0853782401 mos2S003   &Thin1  &  -569.7 &-1577.6& 2019-12-11 &  PrimePartialW3   &21.4   &91.1\%\\
0811012601 mos2S003   &Thin1  &  -182.8 &-1555.1& 2020-02-23 &  PrimePartialW3   &40.2   &89.3\%\\
                      &       &         &      &             &  \textbf{Total}   &813 &
\enddata
\end{deluxetable*}

\clearpage

\bibliography{n132d_xmm}{}

\begin{thebibliography}{}
\expandafter\ifx\csname natexlab\endcsname\relax\def\natexlab#1{#1}\fi
\providecommand{\url}[1]{\href{#1}{#1}}
\providecommand{\dodoi}[1]{doi:~\href{http://doi.org/#1}{\nolinkurl{#1}}}
\providecommand{\doeprint}[1]{\href{http://ascl.net/#1}{\nolinkurl{http://ascl.net/#1}}}
\providecommand{\doarXiv}[1]{\href{https://arxiv.org/abs/#1}{\nolinkurl{https://arxiv.org/abs/#1}}}

\bibitem[{{Arnaud}(1996)}]{1996ASPC..101...17A}
{Arnaud}, K.~A. 1996, in Astronomical Society of the Pacific Conference Series,
  Vol. 101, Astronomical Data Analysis Software and Systems V, ed. G.~H.
  {Jacoby} \& J.~{Barnes}, 17

\bibitem[{{Bamba} {et~al.}(2018){Bamba}, {Ohira}, {Yamazaki}, {Sawada},
  {Terada}, {Koyama}, {Miller}, {Yamaguchi}, {Katsuda}, {Nobukawa}, \&
  {Nobukawa}}]{bamba2018}
{Bamba}, A., {Ohira}, Y., {Yamazaki}, R., {et~al.} 2018, \apj, 854, 71,
  \dodoi{10.3847/1538-4357/aaa5a0}

\bibitem[{{Banas} {et~al.}(1997){Banas}, {Hughes}, {Bronfman}, \&
  {Nyman}}]{banas1997}
{Banas}, K.~R., {Hughes}, J.~P., {Bronfman}, L., \& {Nyman}, L.~{\r{A}}. 1997,
  \apj, 480, 607, \dodoi{10.1086/303989}

\bibitem[{{Banovetz} {et~al.}(2023){Banovetz}, {Milisavljevic}, {Sravan},
  {Weil}, {Subrayan}, {Fesen}, {Patnaude}, {Plucinsky}, {Law}, {Blair}, \&
  {Morse}}]{banovetz2023}
{Banovetz}, J., {Milisavljevic}, D., {Sravan}, N., {et~al.} 2023, \apj, 948,
  33, \dodoi{10.3847/1538-4357/acb8b6}

\bibitem[{{Behar} {et~al.}(2001){Behar}, {Rasmussen}, {Griffiths}, {Dennerl},
  {Audard}, {Aschenbach}, \& {Brinkman}}]{behar2001}
{Behar}, E., {Rasmussen}, A.~P., {Griffiths}, R.~G., {et~al.} 2001, \aap, 365,
  L242, \dodoi{10.1051/0004-6361:20000082}

\bibitem[{{Bethe}(1990)}]{bethe1990}
{Bethe}, H.~A. 1990, Reviews of Modern Physics, 62, 801,
  \dodoi{10.1103/RevModPhys.62.801}

\bibitem[{{Blair} {et~al.}(2000){Blair}, {Morse}, {Raymond}, {Kirshner},
  {Hughes}, {Dopita}, {Sutherland}, {Long}, \& {Winkler}}]{blair2000}
{Blair}, W.~P., {Morse}, J.~A., {Raymond}, J.~C., {et~al.} 2000, \apj, 537,
  667, \dodoi{10.1086/309077}

\bibitem[{{Borkowski} {et~al.}(2007){Borkowski}, {Hendrick}, \&
  {Reynolds}}]{borkowski2007}
{Borkowski}, K.~J., {Hendrick}, S.~P., \& {Reynolds}, S.~P. 2007, \apjl, 671,
  L45, \dodoi{10.1086/524733}

\bibitem[{{Burrows} \& {Vartanyan}(2021)}]{burrows2021}
{Burrows}, A., \& {Vartanyan}, D. 2021, \nat, 589, 29,
  \dodoi{10.1038/s41586-020-03059-w}

\bibitem[{{Cash}(1979)}]{cash1979}
{Cash}, W. 1979, \apj, 228, 939, \dodoi{10.1086/156922}

\bibitem[{{Chen} {et~al.}(2003){Chen}, {Zhang}, {Williams}, \&
  {Wang}}]{chen2003}
{Chen}, Y., {Zhang}, F., {Williams}, R.~M., \& {Wang}, Q.~D. 2003, \apj, 595,
  227, \dodoi{10.1086/377353}

\bibitem[{{Clementini} {et~al.}(2003){Clementini}, {Gratton}, {Bragaglia},
  {Carretta}, {Di Fabrizio}, \& {Maio}}]{2003AJ....125.1309C}
{Clementini}, G., {Gratton}, R., {Bragaglia}, A., {et~al.} 2003, \aj, 125,
  1309, \dodoi{10.1086/367773}

\bibitem[{{Danziger} \& {Dennefeld}(1976)}]{danzinger1976}
{Danziger}, I.~J., \& {Dennefeld}, M. 1976, \apj, 207, 394,
  \dodoi{10.1086/154507}

\bibitem[{{De Luca} \& {Molendi}(2004)}]{2004A&A...419..837D}
{De Luca}, A., \& {Molendi}, S. 2004, \aap, 419, 837,
  \dodoi{10.1051/0004-6361:20034421}

\bibitem[{{den Herder} {et~al.}(2001){den Herder}, {Brinkman}, {Kahn},
  {Branduardi-Raymont}, {Thomsen}, {Aarts}, {Audard}, {Bixler}, {den Boggende},
  {Cottam}, {Decker}, {Dubbeldam}, {Erd}, {Goulooze}, {G{\"u}del}, {Guttridge},
  {Hailey}, {Janabi}, {Kaastra}, {de Korte}, {van Leeuwen}, {Mauche},
  {McCalden}, {Mewe}, {Naber}, {Paerels}, {Peterson}, {Rasmussen}, {Rees},
  {Sakelliou}, {Sako}, {Spodek}, {Stern}, {Tamura}, {Tandy}, {de Vries},
  {Welch}, \& {Zehnder}}]{denHerder2001}
{den Herder}, J.~W., {Brinkman}, A.~C., {Kahn}, S.~M., {et~al.} 2001, \aap,
  365, L7, \dodoi{10.1051/0004-6361:20000058}

\bibitem[{{France} {et~al.}(2009){France}, {Beasley}, {Keeney}, {Danforth},
  {Froning}, {Green}, \& {Shull}}]{2009ApJ...707L..27F}
{France}, K., {Beasley}, M., {Keeney}, B.~A., {et~al.} 2009, \apjl, 707, L27,
  \dodoi{10.1088/0004-637X/707/1/L27}

\bibitem[{{Gabler} {et~al.}(2021){Gabler}, {Wongwathanarat}, \&
  {Janka}}]{gabler2021}
{Gabler}, M., {Wongwathanarat}, A., \& {Janka}, H.-T. 2021, \mnras, 502, 3264,
  \dodoi{10.1093/mnras/stab116}

\bibitem[{{Gordon} \& {Arnaud}(2021)}]{2021ascl.soft01014G}
{Gordon}, C., \& {Arnaud}, K. 2021, {PyXspec: Python interface to XSPEC
  spectral-fitting program}, Astrophysics Source Code Library, record
  ascl:2101.014.
\newblock \doeprint{2101.014}

\bibitem[{{Hammer} {et~al.}(2010){Hammer}, {Janka}, \&
  {M{\"u}ller}}]{hammer2010}
{Hammer}, N.~J., {Janka}, H.~T., \& {M{\"u}ller}, E. 2010, \apj, 714, 1371,
  \dodoi{10.1088/0004-637X/714/2/1371}

\bibitem[{{Heger} {et~al.}(2003){Heger}, {Fryer}, {Woosley}, {Langer}, \&
  {Hartmann}}]{heger2003}
{Heger}, A., {Fryer}, C.~L., {Woosley}, S.~E., {Langer}, N., \& {Hartmann},
  D.~H. 2003, \apj, 591, 288, \dodoi{10.1086/375341}

\bibitem[{{H.E.S.S. Collaboration} {et~al.}(2021){H.E.S.S. Collaboration},
  {Abdalla, H.}, {Aharonian, F.}, {Ait Benkhali, F.}, {Angüner, E. O.},
  {Arcaro, C.}, {Armand, C.}, {Armstrong, T.}, {Ashkar, H.}, {Backes, M.},
  {Baghmanyan, V.}, {Barbosa Martins, V.}, {Barnacka, A.}, {Barnard, M.},
  {Batzofin, R.}, {Becherini, Y.}, {Berge, D.}, {Bernlöhr, K.}, {Bi, B.},
  {Böttcher, M.}, {Boisson, C.}, {Bolmont, J.}, {de Bony de Lavergne, M.},
  {Breuhaus, M.}, {Brose, R.}, {Brun, F.}, {Bulik, T.}, {Bylund, T.}, {Cangemi,
  F.}, {Caroff, S.}, {Casanova, S.}, {Catalano, J.}, {Chambery, P.}, {Chand,
  T.}, {Chen, A.}, {Cotter, G.}, {Curyło, M.}, {Damascene Mbarubucyeye, J.},
  {Davids, I. D.}, {Davies, J.}, {Devin, J.}, {Djannati-Ataï, A.}, {Dmytriiev,
  A.}, {Donath, A.}, {Doroshenko, V.}, {Dreyer, L.}, {Du Plessis, L.}, {Duffy,
  C.}, {Egberts, K.}, {Einecke, S.}, {Ernenwein, J.-P.}, {Fegan, S.}, {Feijen,
  K.}, {Fiasson, A.}, {Fichet de Clairfontaine, G.}, {Fontaine, G.}, {Lott,
  F.}, {Füßling, M.}, {Funk, S.}, {Gabici, S.}, {Gallant, Y. A.}, {Giavitto,
  G.}, {Giunti, L.}, {Glawion, D.}, {Glicenstein, J. F.}, {Grondin, M.-H.},
  {Hattingh, S.}, {Haupt, M.}, {Hermann, G.}, {Hinton, J. A.}, {Hofmann, W.},
  {Hoischen, C.}, {Holch, T. L.}, {Holler, M.}, {Hörbe, M.}, {Horns, D.},
  {Huang, Zhiqiu}, {Huber, D.}, {Jamrozy, M.}, {Jankowsky, F.}, {Joshi, V.},
  {Jung-Richardt, I.}, {Kasai, E.}, {Katarzyński, K.}, {Katz, U.},
  {Khangulyan, D.}, {Khélifi, B.}, {Klepser, S.}, {Kluźniak, W.}, {Komin,
  Nu.}, {Konno, R.}, {Kosack, K.}, {Kostunin, D.}, {Kreter, M.}, {Kukec Mezek,
  G.}, {Kundu, A.}, {Lamanna, G.}, {Le Stum, S.}, {Lemière, A.},
  {Lemoine-Goumard, M.}, {Lenain, J.-P.}, {Leuschner, F.}, {Levy, C.}, {Lohse,
  T.}, {Luashvili, A.}, {Lypova, I.}, {Mackey, J.}, {Majumdar, J.}, {Malyshev,
  D.}, {Malyshev, D.}, {Marandon, V.}, {Marchegiani, P.}, {Marcowith, A.},
  {Mares, A.}, {Martí-Devesa, G.}, {Marx, R.}, {Maurin, G.}, {Meintjes, P.
  J.}, {Meyer, M.}, {Mitchell, A.}, {Moderski, R.}, {Mohrmann, L.}, {Montanari,
  A.}, {Moore, C.}, {Moulin, E.}, {Muller, J.}, {Murach, T.}, {Nakashima, K.},
  {de Naurois, M.}, {Nayerhoda, A.}, {Ndiyavala, H.}, {Niemiec, J.}, {Priyana
  Noel, A.}, {O’Brien, P.}, {Oberholzer, L.}, {Odaka, H.}, {Ohm, S.},
  {Olivera-Nieto, L.}, {de Ona Wilhelmi, E.}, {Ostrowski, M.}, {Panny, S.},
  {Panter, M.}, {Parsons, R. D.}, {Peron, G.}, {Pita, S.}, {Poireau, V.},
  {Prokhorov, D. A.}, {Prokoph, H.}, {Pühlhofer, G.}, {Punch, M.},
  {Quirrenbach, A.}, {Reichherzer, P.}, {Reimer, A.}, {Reimer, O.}, {Remy, Q.},
  {Renaud, M.}, {Reville, B.}, {Rieger, F.}, {Romoli, C.}, {Rowell, G.},
  {Rudak, B.}, {Rueda Ricarte, H.}, {Ruiz-Velasco, E.}, {Sahakian, V.},
  {Sailer, S.}, {Salzmann, H.}, {Sanchez, D. A.}, {Santangelo, A.}, {Sasaki,
  M.}, {Schäfer, J.}, {Schüssler, F.}, {Schutte, H. M.}, {Schwanke, U.},
  {Senniappan, M.}, {Seyffert, A. S.}, {Shapopi, J. N. S.}, {Shiningayamwe,
  K.}, {Simoni, R.}, {Sinha, A.}, {Sol, H.}, {Specovius, A.}, {Spencer, S.},
  {Spir-Jacob, M.}, {Stawarz, Ł.}, {Steenkamp, R.}, {Stegmann, C.},
  {Steinmassl, S.}, {Steppa, C.}, {Sun, L.}, {Takahashi, T.}, {Tanaka, T.},
  {Tavernier, T.}, {Taylor, A. M.}, {Terrier, R.}, {Thiersen, J. H. E.},
  {Thorpe-Morgan, C.}, {Tluczykont, M.}, {Tomankova, L.}, {Tsirou, M.}, {Tsuji,
  N.}, {Tuffs, R.}, {Uchiyama, Y.}, {van der Walt, D. J.}, {van Eldik, C.},
  {van Rensburg, C.}, {van Soelen, B.}, {Vasileiadis, G.}, {Veh, J.}, {Venter,
  C.}, {Vincent, P.}, {Vink, J.}, {Völk, H. J.}, {Wagner, S. J.}, {Watson,
  J.}, {Werner, F.}, {White, R.}, {Wierzcholska, A.}, {Wong, Yu Wun}, {Yassin,
  H.}, {Yusafzai, A.}, {Zacharias, M.}, {Zanin, R.}, {Zargaryan, D.},
  {Zdziarski, A. A.}, {Zech, A.}, {Zhu, S. J.}, {Zmija, A.}, {Zouari, S.}, \&
  {Żywucka, N.}}]{2021arXiv210802015H}
{H.E.S.S. Collaboration}, {Abdalla, H.}, {Aharonian, F.}, {et~al.} 2021, \aap,
  655, A7, \dodoi{10.1051/0004-6361/202141486}

\bibitem[{{Hitomi Collaboration} {et~al.}(2018){Hitomi Collaboration},
  {Aharonian}, {Akamatsu}, {Akimoto}, {Allen}, {Angelini}, {Audard}, {Awaki},
  {Axelsson}, {Bamba}, {Bautz}, {Blandford}, {Brenneman}, {Brown}, {Bulbul},
  {Cackett}, {Chernyakova}, {Chiao}, {Coppi}, {Costantini}, {de Plaa}, {de
  Vries}, {den Herder}, {Done}, {Dotani}, {Ebisawa}, {Eckart}, {Enoto}, {Ezoe},
  {Fabian}, {Ferrigno}, {Foster}, {Fujimoto}, {Fukazawa}, {Furuzawa},
  {Galeazzi}, {Gallo}, {Gandhi}, {Giustini}, {Goldwurm}, {Gu}, {Guainazzi},
  {Haba}, {Hagino}, {Hamaguchi}, {Harrus}, {Hatsukade}, {Hayashi}, {Hayashi},
  {Hayashida}, {Hiraga}, {Hornschemeier}, {Hoshino}, {Hughes}, {Ichinohe},
  {Iizuka}, {Inoue}, {Inoue}, {Ishida}, {Ishikawa}, {Ishisaki}, {Iwai},
  {Kaastra}, {Kallman}, {Kamae}, {Kataoka}, {Katsuda}, {Kawai}, {Kelley},
  {Kilbourne}, {Kitaguchi}, {Kitamoto}, {Kitayama}, {Kohmura}, {Kokubun},
  {Koyama}, {Koyama}, {Kretschmar}, {Krimm}, {Kubota}, {Kunieda}, {Laurent},
  {Lee}, {Leutenegger}, {Limousin}, {Loewenstein}, {Long}, {Lumb}, {Madejski},
  {Maeda}, {Maier}, {Makishima}, {Markevitch}, {Matsumoto}, {Matsushita},
  {McCammon}, {McNamara}, {Mehdipour}, {Miller}, {Miller}, {Mineshige},
  {Mitsuda}, {Mitsuishi}, {Miyazawa}, {Mizuno}, {Mori}, {Mori}, {Mukai},
  {Murakami}, {Mushotzky}, {Nakagawa}, {Nakajima}, {Nakamori}, {Nakashima},
  {Nakazawa}, {Nobukawa}, {Nobukawa}, {Noda}, {Odaka}, {Ohashi}, {Ohno},
  {Okajima}, {Ota}, {Ozaki}, {Paerels}, {Paltani}, {Petre}, {Pinto}, {Porter},
  {Pottschmidt}, {Reynolds}, {Safi-Harb}, {Saito}, {Sakai}, {Sasaki}, {Sato},
  {Sato}, {Sato}, {Sato}, {Sawada}, {Schartel}, {Serlemtsos}, {Seta},
  {Shidatsu}, {Simionescu}, {Smith}, {Soong}, {Stawarz}, {Sugawara}, {Sugita},
  {Szymkowiak}, {Tajima}, {Takahashi}, {Takahashi}, {Takeda}, {Takei},
  {Tamagawa}, {Tamura}, {Tanaka}, {Tanaka}, {Tanaka}, {Tashiro}, {Tawara},
  {Terada}, {Terashima}, {Tombesi}, {Tomida}, {Tsuboi}, {Tsujimoto}, {Tsunemi},
  {Tsuru}, {Uchida}, {Uchiyama}, {Uchiyama}, {Ueda}, {Ueda}, {Uno}, {Urry},
  {Ursino}, {Watanabe}, {Werner}, {Wilkins}, {Williams}, {Yamada}, {Yamaguchi},
  {Yamaoka}, {Yamasaki}, {Yamauchi}, {Yamauchi}, {Yaqoob}, {Yatsu}, {Yonetoku},
  {Zhuravleva}, \& {Zoghbi}}]{hitomi2018}
{Hitomi Collaboration}, {Aharonian}, F., {Akamatsu}, H., {et~al.} 2018, \pasj,
  70, 16, \dodoi{10.1093/pasj/psx151}

\bibitem[{{Hughes}(1987)}]{hughes1987}
{Hughes}, J.~P. 1987, \apj, 314, 103, \dodoi{10.1086/165043}

\bibitem[{{Hughes} {et~al.}(2000){Hughes}, {Rakowski}, {Burrows}, \&
  {Slane}}]{hughes2000}
{Hughes}, J.~P., {Rakowski}, C.~E., {Burrows}, D.~N., \& {Slane}, P.~O. 2000,
  \apjl, 528, L109, \dodoi{10.1086/312438}

\bibitem[{{Hwang} \& {Laming}(2012)}]{hwang2012}
{Hwang}, U., \& {Laming}, J.~M. 2012, \apj, 746, 130,
  \dodoi{10.1088/0004-637X/746/2/130}

\bibitem[{{Janka}(2017)}]{janka2017J}
{Janka}, H.-T. 2017, {Neutrino-Driven Explosions}, ed. A.~W. {Alsabti} \&
  P.~{Murdin} (Springer Cham), 1095, \dodoi{10.1007/978-3-319-21846-5\_109}

\bibitem[{{Katsuda} {et~al.}(2018){Katsuda}, {Takiwaki}, {Tominaga}, {Moriya},
  \& {Nakamura}}]{2018ApJ...863..127K}
{Katsuda}, S., {Takiwaki}, T., {Tominaga}, N., {Moriya}, T.~J., \& {Nakamura},
  K. 2018, \apj, 863, 127, \dodoi{10.3847/1538-4357/aad2d8}

\bibitem[{{Kelley} {et~al.}(2016){Kelley}, {Akamatsu}, {Azzarello}, {Bialas},
  {Boyce}, {Brown}, {Canavan}, {Chiao}, {Costantini}, {DiPirro}, {Eckart},
  {Ezoe}, {Fujimoto}, {Haas}, {den Herder}, {Hoshino}, {Ishikawa}, {Ishisaki},
  {Iyomoto}, {Kilbourne}, {Kimball}, {Kitamoto}, {Konami}, {Koyama},
  {Leutenegger}, {McCammon}, {Mitsuda}, {Mitsuishi}, {Moseley}, {Murakami},
  {Murakami}, {Noda}, {Ogawa}, {Ohashi}, {Okamoto}, {Ota}, {Paltani}, {Porter},
  {Sakai}, {Sato}, {Sato}, {Sawada}, {Seta}, {Shinozaki}, {Shirron},
  {Sneiderman}, {Sugita}, {Szymkowiak}, {Takei}, {Tamagawa}, {Tashiro},
  {Terada}, {Tsujimoto}, {de Vries}, {Yamada}, {Yamasaki}, \&
  {Yatsu}}]{kelley2016}
{Kelley}, R.~L., {Akamatsu}, H., {Azzarello}, P., {et~al.} 2016, in Society of
  Photo-Optical Instrumentation Engineers (SPIE) Conference Series, Vol. 9905,
  Space Telescopes and Instrumentation 2016: Ultraviolet to Gamma Ray, ed.
  J.-W.~A. {den Herder}, T.~{Takahashi}, \& M.~{Bautz}, 99050V,
  \dodoi{10.1117/12.2232509}

\bibitem[{{Lasker}(1978)}]{lasker1978}
{Lasker}, B.~M. 1978, \apj, 223, 109, \dodoi{10.1086/156241}

\bibitem[{{Lasker}(1980)}]{lasker1980}
---. 1980, \apj, 237, 765, \dodoi{10.1086/157923}

\bibitem[{{Law} {et~al.}(2020){Law}, {Milisavljevic}, {Patnaude}, {Plucinsky},
  {Gladders}, {Schmidt}, {Sravan}, {Banovetz}, {Sano}, {McGraw}, {Takahashi},
  \& {Orlando}}]{law2020}
{Law}, C.~J., {Milisavljevic}, D., {Patnaude}, D.~J., {et~al.} 2020, \apj, 894,
  73, \dodoi{10.3847/1538-4357/ab873a}

\bibitem[{{Lazendic} {et~al.}(2006){Lazendic}, {Dewey}, {Schulz}, \&
  {Canizares}}]{lazendic2006}
{Lazendic}, J.~S., {Dewey}, D., {Schulz}, N.~S., \& {Canizares}, C.~R. 2006,
  \apj, 651, 250, \dodoi{10.1086/507481}

\bibitem[{{Maggi} {et~al.}(2016){Maggi}, {Haberl}, {Kavanagh}, {Sasaki},
  {Bozzetto}, {Filipovi{\'c}}, {Vasilopoulos}, {Pietsch}, {Points}, {Chu},
  {Dickel}, {Ehle}, {Williams}, \& {Greiner}}]{maggi2016}
{Maggi}, P., {Haberl}, F., {Kavanagh}, P.~J., {et~al.} 2016, \aap, 585, A162,
  \dodoi{10.1051/0004-6361/201526932}

\bibitem[{{Morse} {et~al.}(1995){Morse}, {Winkler}, \& {Kirshner}}]{morse1995}
{Morse}, J.~A., {Winkler}, P.~F., \& {Kirshner}, R.~P. 1995, \aj, 109, 2104,
  \dodoi{10.1086/117436}

\bibitem[{{Orlando} {et~al.}(2015){Orlando}, {Miceli}, {Pumo}, \&
  {Bocchino}}]{orlando2015}
{Orlando}, S., {Miceli}, M., {Pumo}, M.~L., \& {Bocchino}, F. 2015, \apj, 810,
  168, \dodoi{10.1088/0004-637X/810/2/168}

\bibitem[{{Orlando} {et~al.}(2016){Orlando}, {Miceli}, {Pumo}, \&
  {Bocchino}}]{orlando2016}
---. 2016, \apj, 822, 22, \dodoi{10.3847/0004-637X/822/1/22}

\bibitem[{{Orlando} {et~al.}(2021){Orlando}, {Wongwathanarat}, {Janka},
  {Miceli}, {Ono}, {Nagataki}, {Bocchino}, \& {Peres}}]{orlando2021}
{Orlando}, S., {Wongwathanarat}, A., {Janka}, H.~T., {et~al.} 2021, \aap, 645,
  A66, \dodoi{10.1051/0004-6361/202039335}

\bibitem[{{Orlando} {et~al.}(2020){Orlando}, {Ono}, {Nagataki}, {Miceli},
  {Umeda}, {Ferrand}, {Bocchino}, {Petruk}, {Peres}, {Takahashi}, \&
  {Yoshida}}]{orlando2020}
{Orlando}, S., {Ono}, M., {Nagataki}, S., {et~al.} 2020, \aap, 636, A22,
  \dodoi{10.1051/0004-6361/201936718}

\bibitem[{{Pietrzy{\'n}ski} {et~al.}(2013){Pietrzy{\'n}ski}, {Graczyk},
  {Gieren}, {Thompson}, {Pilecki}, {Udalski}, {Soszy{\'n}ski}, {Koz{\l}owski},
  {Konorski}, {Suchomska}, {Bono}, {Moroni}, {Villanova}, {Nardetto},
  {Bresolin}, {Kudritzki}, {Storm}, {Gallenne}, {Smolec}, {Minniti}, {Kubiak},
  {Szyma{\'n}ski}, {Poleski}, {Wyrzykowski}, {Ulaczyk}, {Pietrukowicz},
  {G{\'o}rski}, \& {Karczmarek}}]{2013Natur.495...76P}
{Pietrzy{\'n}ski}, G., {Graczyk}, D., {Gieren}, W., {et~al.} 2013, \nat, 495,
  76, \dodoi{10.1038/nature11878}

\bibitem[{{Pietrzy{\'n}ski} {et~al.}(2019){Pietrzy{\'n}ski}, {Graczyk},
  {Gallenne}, {Gieren}, {Thompson}, {Pilecki}, {Karczmarek}, {G{\'o}rski},
  {Suchomska}, {Taormina}, {Zgirski}, {Wielg{\'o}rski}, {Ko{\l}aczkowski},
  {Konorski}, {Villanova}, {Nardetto}, {Kervella}, {Bresolin}, {Kudritzki},
  {Storm}, {Smolec}, \& {Narloch}}]{2019Natur.567..200P}
{Pietrzy{\'n}ski}, G., {Graczyk}, D., {Gallenne}, A., {et~al.} 2019, \nat, 567,
  200, \dodoi{10.1038/s41586-019-0999-4}

\bibitem[{{Ravi} {et~al.}(2021){Ravi}, {Park}, {Zhekov}, {Miceli}, {Orlando},
  {Frank}, \& {Burrows}}]{ravi2021}
{Ravi}, A.~P., {Park}, S., {Zhekov}, S.~A., {et~al.} 2021, \apj, 922, 140,
  \dodoi{10.3847/1538-4357/ac249a}

\bibitem[{{Russell} \& {Dopita}(1992)}]{1992ApJ...384..508R}
{Russell}, S.~C., \& {Dopita}, M.~A. 1992, \apj, 384, 508,
  \dodoi{10.1086/170893}

\bibitem[{{Rutherford} {et~al.}(2013){Rutherford}, {Dewey},
  {Figueroa-Feliciano}, {Heine}, {Bastien}, {Sato}, \&
  {Canizares}}]{rutherford2013}
{Rutherford}, J., {Dewey}, D., {Figueroa-Feliciano}, E., {et~al.} 2013, \apj,
  769, 64, \dodoi{10.1088/0004-637X/769/1/64}

\bibitem[{{Sano} {et~al.}(2017){Sano}, {Fujii}, {Yamane}, {Inaba}, {Yoshiike},
  {Fukuda}, {Voisin}, {Rowell}, \& {Fukui}}]{sano2017}
{Sano}, H., {Fujii}, K., {Yamane}, Y., {et~al.} 2017, in American Institute of
  Physics Conference Series, Vol. 1792, 6th International Symposium on High
  Energy Gamma-Ray Astronomy, 040038, \dodoi{10.1063/1.4968942}

\bibitem[{{Sano} {et~al.}(2020){Sano}, {Plucinsky}, {Bamba}, {Sharda},
  {Filipovi{\'c}}, {Law}, {Alsaberi}, {Yamane}, {Tokuda}, {Acero}, {Sasaki},
  {Vink}, {Inoue}, {Inutsuka}, {Shimoda}, {Tsuge}, {Fujii}, {Voisin}, {Maxted},
  {Rowell}, {Onishi}, {Kawamura}, {Mizuno}, {Yamamoto}, {Tachihara}, \&
  {Fukui}}]{sano2020}
{Sano}, H., {Plucinsky}, P.~P., {Bamba}, A., {et~al.} 2020, \apj, 902, 53,
  \dodoi{10.3847/1538-4357/abb469}

\bibitem[{{Sapienza} {et~al.}(2024){Sapienza}, {Miceli}, {Bamba}, {Orlando},
  {Lee}, {Nagataki}, {Ono}, {Katsuda}, {Mori}, {Sawada}, {Terada}, {Giuffrida},
  \& {Bocchino}}]{sapienza2023}
{Sapienza}, V., {Miceli}, M., {Bamba}, A., {et~al.} 2024, \apjl, 961, L9,
  \dodoi{10.3847/2041-8213/ad16e3}

\bibitem[{{Sato} {et~al.}(2021){Sato}, {Maeda}, {Nagataki}, {Yoshida},
  {Grefenstette}, {Williams}, {Umeda}, {Ono}, \& {Hughes}}]{sato2021}
{Sato}, T., {Maeda}, K., {Nagataki}, S., {et~al.} 2021, \nat, 592, 537,
  \dodoi{10.1038/s41586-021-03391-9}

\bibitem[{{Schenck} {et~al.}(2016){Schenck}, {Park}, \&
  {Post}}]{2016AJ....151..161S}
{Schenck}, A., {Park}, S., \& {Post}, S. 2016, \aj, 151, 161,
  \dodoi{10.3847/0004-6256/151/6/161}

\bibitem[{{Sharda} {et~al.}(2020){Sharda}, {Gaetz}, {Kashyap}, \&
  {Plucinsky}}]{sharda2020}
{Sharda}, P., {Gaetz}, T.~J., {Kashyap}, V.~L., \& {Plucinsky}, P.~P. 2020,
  \apj, 894, 145, \dodoi{10.3847/1538-4357/ab8a46}

\bibitem[{{Siegel} {et~al.}(2021){Siegel}, {Dwarkadas}, {Frank}, \&
  {Burrows}}]{siegel2021}
{Siegel}, J., {Dwarkadas}, V.~V., {Frank}, K.~A., \& {Burrows}, D.~N. 2021,
  \apj, 922, 67, \dodoi{10.3847/1538-4357/ac2305}

\bibitem[{{Sukhbold} {et~al.}(2016){Sukhbold}, {Ertl}, {Woosley}, {Brown}, \&
  {Janka}}]{2016ApJ...821...38S}
{Sukhbold}, T., {Ertl}, T., {Woosley}, S.~E., {Brown}, J.~M., \& {Janka}, H.~T.
  2016, \apj, 821, 38, \dodoi{10.3847/0004-637X/821/1/38}

\bibitem[{{Sun} {et~al.}(2025){Sun}, {Orlando}, {Greco}, {Miceli}, {Chen},
  {Vink}, \& {Zhou}}]{sun2025}
{Sun}, L., {Orlando}, S., {Greco}, E., {et~al.} 2025, \apj, 981, 26,
  \dodoi{10.3847/1538-4357/adaea7}

\bibitem[{{Suzuki} {et~al.}(2020){Suzuki}, {Yamaguchi}, {Ishida}, {Uchida},
  {Plucinsky}, {Foster}, \& {Miller}}]{suzuki2020}
{Suzuki}, H., {Yamaguchi}, H., {Ishida}, M., {et~al.} 2020, \apj, 900, 39,
  \dodoi{10.3847/1538-4357/aba524}

\bibitem[{{Vink}(2017)}]{vink2017}
{Vink}, J. 2017, {X-Ray Emission Properties of Supernova Remnants}, ed. A.~W.
  {Alsabti} \& P.~{Murdin} (Springer Cham), 2063,
  \dodoi{10.1007/978-3-319-21846-5\_92}

\bibitem[{{Vogt} \& {Dopita}(2011)}]{vogt2011}
{Vogt}, F., \& {Dopita}, M.~A. 2011, \apss, 331, 521,
  \dodoi{10.1007/s10509-010-0479-7}

\bibitem[{Wilms {et~al.}(2000)Wilms, Allen, \& McCray}]{Wilms2000}
Wilms, J., Allen, A., \& McCray, R. 2000, The Astrophysical Journal, 542, 914,
  \dodoi{10.1086/317016}

\bibitem[{{Wongwathanarat} {et~al.}(2015){Wongwathanarat}, {M{\"u}ller}, \&
  {Janka}}]{wongwathanarat2015}
{Wongwathanarat}, A., {M{\"u}ller}, E., \& {Janka}, H.~T. 2015, \aap, 577, A48,
  \dodoi{10.1051/0004-6361/201425025}

\bibitem[{{XRISM Collaboration} {et~al.}(2024){XRISM Collaboration}, {Audard},
  {Awaki}, {Ballhausen}, {Bamba}, {Behar}, {Boissay-Malaquin}, {Brenneman},
  {Brown}, {Corrales}, {Costantini}, {Cumbee}, {Diaz-Trigo}, {Done}, {Dotani},
  {Ebisawa}, {Eckart}, {Eckert}, {Enoto}, {Eguchi}, {Ezoe}, {Foster},
  {Fujimoto}, {Fujita}, {Fukazawa}, {Fukushima}, {Furuzawa}, {Gallo},
  {Garc{\'\i}a}, {Gu}, {Guainazzi}, {Hagino}, {Hamaguchi}, {Hatsukade},
  {Hayashi}, {Hayashi}, {Hell}, {Hodges-Kluck}, {Hornschemeier}, {Ichinohe},
  {Ishida}, {Ishikawa}, {Ishisaki}, {Kaastra}, {Kallman}, {Kara}, {Katsuda},
  {Kanemaru}, {Kelley}, {Kilbourne}, {Kitamoto}, {Kobayashi}, {Kohmura},
  {Kubota}, {Leutenegger}, {Loewenstein}, {Maeda}, {Markevitch}, {Matsumoto},
  {Matsushita}, {McCammon}, {McNamara}, {Mernier}, {Miller}, {Miller},
  {Mitsuishi}, {Mizumoto}, {Mizuno}, {Mori}, {Mukai}, {Murakami}, {Mushotzky},
  {Nakajima}, {Nakazawa}, {Ness}, {Nobukawa}, {Nobukawa}, {Noda}, {Odaka},
  {Ogawa}, {Ogorzalek}, {Okajima}, {Ota}, {Paltani}, {Petre}, {Plucinsky},
  {Porter}, {Pottschmidt}, {Sato}, {Sato}, {Sawada}, {Seta}, {Shidatsu},
  {Simionescu}, {Smith}, {Suzuki}, {Szymkowiak}, {Takahashi}, {Takeo},
  {Tamagawa}, {Tamura}, {Tanaka}, {Tanimoto}, {Tashiro}, {Terada}, {Terashima},
  {Tsuboi}, {Tsujimoto}, {Tsunemi}, {Tsuru}, {Uchida}, {Uchida}, {Uchida},
  {Uchiyama}, {Ueda}, {Uno}, {Vink}, {Watanabe}, {Williams}, {Yamada},
  {Yamada}, {Yamaguchi}, {Yamaoka}, {Yamasaki}, {Yamauchi}, {Yamauchi},
  {Yaqoob}, {Yoneyama}, {Yoshida}, {Yukita}, {Zhuravleva}, {Agarwal}, \&
  {Ohshiro}}]{2024_XRISM_N132D}
{XRISM Collaboration}, {Audard}, M., {Awaki}, H., {et~al.} 2024, \pasj, 76,
  1186, \dodoi{10.1093/pasj/psae080}

\bibitem[{{Yamaguchi} {et~al.}(2014{\natexlab{a}}){Yamaguchi}, {Badenes},
  {Petre}, {Nakano}, {Castro}, {Enoto}, {Hiraga}, {Hughes}, {Maeda},
  {Nobukawa}, {Safi-Harb}, {Slane}, {Smith}, \& {Uchida}}]{yamaguchi2014}
{Yamaguchi}, H., {Badenes}, C., {Petre}, R., {et~al.} 2014{\natexlab{a}},
  \apjl, 785, L27, \dodoi{10.1088/2041-8205/785/2/L27}

\bibitem[{{Yamaguchi} {et~al.}(2014{\natexlab{b}}){Yamaguchi}, {Eriksen},
  {Badenes}, {Hughes}, {Brickhouse}, {Foster}, {Patnaude}, {Petre}, {Slane}, \&
  {Smith}}]{2014ApJ...780..136Y}
{Yamaguchi}, H., {Eriksen}, K.~A., {Badenes}, C., {et~al.} 2014{\natexlab{b}},
  \apj, 780, 136, \dodoi{10.1088/0004-637X/780/2/136}

\bibitem[{Yerokhin \& Surzhykov(2019)}]{10.1063/1.5121413}
Yerokhin, V.~A., \& Surzhykov, A. 2019, Journal of Physical and Chemical
  Reference Data, 48, 033104, \dodoi{10.1063/1.5121413}

\end{thebibliography}
\bibliographystyle{aasjournal}

\end{document}